\documentclass[fleqn,usenatbib]{mnras}

\usepackage{newtxtext,newtxmath}

\usepackage[T1]{fontenc}

\DeclareRobustCommand{\VAN}[3]{#2}
\let\VANthebibliography\thebibliography
\def\thebibliography{\DeclareRobustCommand{\VAN}[3]{##3}\VANthebibliography}

\usepackage{graphicx}	
\usepackage{amsmath}	
\usepackage{tikz}
\usetikzlibrary{svg.path}
\usepackage{scalerel}

\usepackage{color}
\usepackage{ulem}

\newcommand{\add}[1]{\textcolor{black}{#1}}

\definecolor{orcidlogocol}{HTML}{A6CE39}
\tikzset{
  orcidlogo/.pic={
    \fill[orcidlogocol] svg{M256,128c0,70.7-57.3,128-128,128C57.3,256,0,198.7,0,128C0,57.3,57.3,0,128,0C198.7,0,256,57.3,256,128z};
    \fill[white] svg{M86.3,186.2H70.9V79.1h15.4v48.4V186.2z}
                 svg{M108.9,79.1h41.6c39.6,0,57,28.3,57,53.6c0,27.5-21.5,53.6-56.8,53.6h-41.8V79.1z M124.3,172.4h24.5c34.9,0,42.9-26.5,42.9-39.7c0-21.5-13.7-39.7-43.7-39.7h-23.7V172.4z}
                 svg{M88.7,56.8c0,5.5-4.5,10.1-10.1,10.1c-5.6,0-10.1-4.6-10.1-10.1c0-5.6,4.5-10.1,10.1-10.1C84.2,46.7,88.7,51.3,88.7,56.8z};
  }
}

\newcommand\orcidicon[1]{\href{https://orcid.org/#1}{\mbox{\scalerel*{
\begin{tikzpicture}[yscale=-1,transform shape]
\pic{orcidlogo};
\end{tikzpicture}
}{|}}}}

\usepackage{hyperref}

\title[Nodal Precession of WASP-33b]{Nodal Precession of WASP-33b for Eleven Years by Doppler Tomographic and Transit Photometric Observations}

\author[N. Watanabe et al.]{
Noriharu Watanabe \orcidicon{0000-0002-7522-8195},$^{1}$\thanks{E-mail: n-watanabe@g.ecc.u-tokyo.ac.jp} 
Norio Narita \orcidicon{0000-0001-8511-2981},$^{2,3,4}$
Enric Palle \orcidicon{0000-0003-0987-1593},$^{4,5}$
Akihiko Fukui \orcidicon{0000-0002-4909-5763},$^{2,4}$
Nobuhiko Kusakabe \orcidicon{0000-0001-9194-1268},$^{3,6}$
\newauthor{
Hannu Parviainen \orcidicon{0000-0001-5519-1391},$^{4,5}$
Felipe Murgas \orcidicon{0000-0001-9087-1245},$^{4,5}$
N\'{u}ria Casasayas-Barris \orcidicon{0000-0002-2891-8222},$^{7}$
Marshall C. Johnson \orcidicon{0000-0002-5099-8185},$^{8}$}
\newauthor{
Bun'ei Sato \orcidicon{0000-0001-8033-5633},$^{9}$
John H. Livingston \orcidicon{0000-0002-4881-3620},$^{10}$
Jerome P. de Leon \orcidicon{0000-0002-6424-3410},$^{10}$
Mayuko Mori \orcidicon{0000-0003-1368-6593},$^{10}$
Taku Nishiumi \orcidicon{0000-0003-1510-8981},$^{6,11}$}
\newauthor{
Yuka Terada \orcidicon{0000-0003-2887-6381},$^{12,13}$
Emma Esparza-Borges \orcidicon{0000-0002-2341-3233},$^{4,5}$
and
Kiyoe Kawauchi \orcidicon{0000-0003-1205-5108}$^{4}$}
\\ \\
$^{1}$Department of Multi-Disciplinary Sciences, Graduate School of Arts and Sciences, The University of Tokyo, 3-8-1 Komaba, Meguro, Tokyo 153-8902, Japan \\
$^{2}$Komaba Institute for Science, The University of Tokyo, 3-8-1 Komaba, Meguro, Tokyo 153-8902, Japan\\
$^{3}$Astrobiology Center, 2-21-1 Osawa, Mitaka, Tokyo 181-8588, Japan\\
$^{4}$Instituto de Astrof\'isica de Canarias (IAC), E-38200 La Laguna, Tenerife, Spain\\
$^{5}$Dept. Astrof\'isica, Universidad de La Laguna (ULL), E-38206 La Laguna, Tenerife, Spain\\
$^{6}$National Astronomical Observatory of Japan, 2-21-1 Osawa, Mitaka, Tokyo 181-8588, Japan\\
$^{7}$Leiden Observatory, Leiden University, Postbus 9513, 2300RA Leiden, The Netherlands\\
$^{8}$Department of Astronomy, The Ohio State University, 4055 McPherson Laboratory, 140 West 18th Ave., Columbus, OH 43210 USA\\
$^{9}$Department of Earth and Planetary Sciences, Tokyo Institute of Technology, Meguro-ku, Tokyo, 152-8551, Japan\\
$^{10}$Department of Astronomy, Graduate School of Science, The University of Tokyo, 7-3-1 Hongo, Bunkyo-ku, Tokyo 113-0033, Japan\\
$^{11}$Department of Astronomical Science, The Graduated University for Advanced Studies, SOKENDAI, 2-21-1, Osawa, Mitaka, Tokyo, 181-8588, Japan\\
$^{12}$Institute of Astronomy and Astrophysics, Academia Sinica, P.O. Box 23-141, Taipei 10617, Taiwan, R.O.C.\\
$^{13}$Department of Astrophysics, National Taiwan University, Taipei 10617, Taiwan, R.O.C.}

\date{Accepted XXX. Received YYY; in original form ZZZ}

\pubyear{2020}

\begin{document}
\label{firstpage}
\pagerange{\pageref{firstpage}--\pageref{lastpage}}
\maketitle

\begin{abstract}
WASP-33b, a hot Jupiter around a hot star, is a rare system in which nodal precession has been discovered. We updated the model for the nodal precession of WASP-33b by adding new observational points. Consequently, we found a motion of the nodal precession spanning 11 years. We present homogenous Doppler tomographic analyses of eight datasets, including two new datasets from TS23 and HIDES, obtained between 2008 and 2019, to illustrate the variations in the projected spin-orbit obliquity of WASP-33b and its impact parameter. We also present its impact parameters based on photometric transit observations captured by MuSCAT in 2017 and MuSCAT2 in 2018. We derived its real spin-orbit obliquity $\psi$, stellar spin inclination $i_{s}$, and stellar gravitational quadrupole moment $J_2$ from the time variation models of the two orbital parameters. We obtained $\psi = 108.19^{+0.95}_{-0.97}$ deg, $i_s = 58.3^{+4.6}_{-4.2}$ deg, and $J_2=(1.36^{+0.15}_{-0.12}) \times 10^{-4}$. Our $J_2$ value was slightly smaller than the theoretically predicted value, which may indicate that its actual stellar internal structure is different from the theoretical one. We derived the nodal precession speed $\dot{\theta}=0.507^{+0.025}_{-0.022}$ deg year$^{-1}$, and its period $P_{\mathrm{pre}}=709^{+33}_{-34}$ years, and found that WASP-33b transits in front of WASP-33 for only $\sim$ 20 \% of the entire nodal precession period.
\end{abstract}




\section{Introduction} \label{sec:intro}
\add{To date,} WASP \citep[Wide Angle Search for Planets;][]{2007MNRAS.375..951C}, KELT \citep[Kilodegree Extremely Little Telescope;][]{2007PASP..119..923P} and TESS \citep{2015JATIS...1a4003R} surveys \add{have} confirm\add{ed} that there are 17 hot Jupiters around hot stars ($>7000$ K). Despite their small number, their orbital obliquities tend to have a wide range. This tendency indicates that they did not \add{experience} realignment by the tidal torque \citep{2012ApJ...757...18A}. \add{Generally}, hot host stars rotate rapidly, \add{yielding} oblateness larger than those of slowly rotating stars. A misaligned orbit and a fast-rotating star force faster \add{precession}.\add{At present}, there are only two planets whose nodal precessions have been detected: Kepler-13Ab \citep{2012MNRAS.421L.122S} and WASP-33b \citep{2015ApJ...810L..23J}. These planets are hot Jupiters\add{,} revolving in misaligned orbits around rapidly rotating hot stars.

WASP-33b, a hot Jupiter ($R_p=1.5 R_J$) around an A-type ($T_{\mathrm{eff}}=7430 \pm 100$K) and rapidly rotating ($V\sin i_s = 85.6$ km s$^{-1}$) star, is the only planet whose precession has been detected by Doppler tomography based on transit spectral data. This planet was discovered by \citet{2010MNRAS.407..507C}\add{;} its nodal precession has been detected by more than one Doppler tomographic measurement \citep{2015ApJ...810L..23J, 2016MNRAS.455..207I, 2020PASJ...72...19W, 2021A&A...653A.104B}. The latest study on this topic \citep{2021A&A...653A.104B} derived the angle between the stellar spin axis and line of sight $i_s = 90.11 \pm 0.12$ deg as well as the stellar gravitational quadrupole moment $J_2 = (6.73 \pm 0.22) \times 10^{-5}$ from the measurements; however, the value of $i_s$ does not match that of \cite{2016MNRAS.455..207I} within 3 $\sigma$, \add{whereas} the value of $J_2$ is \add{also} not consistent with that of \cite{2020PASJ...72...19W} within 3 $\sigma$. Moreover, \citep{2021A&A...653A.104B} revealed that the change \add{in the transit chord} of WASP-33b is \add{slightly overly} complicated \add{for hereafter estimating} the position of the transit chord. Thus, we performed more additional observations and datasets to \add{obtain} more accurate values of $i_s$ and $J_2$,\add{as well as a} clearer forecasting of the transit chord. 

In this paper, we report \add{the nodal precession of} WASP-33b for a longer period of 11 years by considering additional Doppler tomographic measurements obtained through spectral transit and transit photometric observations. Transit photometry is an important method for measuring the impact parameter $b$ and its change, as \add{demonstrated by} \citet{2012MNRAS.421L.122S}\add{, who} applied \add{this} to detect \add{the nodal precession of} Kepler-13Ab using Kepler's photometric datasets. In Section \ref{sec:DT}, we present our spectral datasets for Doppler tomography and the steps for measuring the orbital parameters, \add{i.e.,} $b$ and the projected spin-orbit obliquity $\lambda$, the angle between the stellar spin axis and the planetary orbital axis. We explain how to handle our transit photometric datasets for determining $b$ in Section \ref{sec:PH}. Moreover, we fit the values of these two parameters from the spectral and photometric measurements with the nodal precession model, which is described in Section \ref{FitMod}. We display the behaviour of WASP-33b's nodal precession and the derived parameters of WASP-33b and its host star in Section \ref{result}. In Section \ref{Discuss}, we \add{discuss} the nodal precession \add{results}. Finally, we present our conclusions in Section \ref{Concl}.

\section{Doppler Tomographic Measurement} \label{sec:DT}
We can \add{simultaneously} measure $\lambda$ and $b$ from the transit spectral data via Doppler tomography. When a planet passes in front of the stellar disk, a bump\add{, referred to as} a planetary shadow\add{,} appears in the stellar line profile. The orbital configuration of the planet can be derived from this shadow motion. 

\subsection{Observation Datasets} \label{sec:obs}
We used eight spectroscopic datasets \add{for} WASP-33 around planetary transits.

One of them was obtained with the High Dispersion Spectrograph \citep[HDS;][]{10.1093/pasj/54.6.855} at the 8.2 m Subaru telescope on 19 October, 2011 UT. The other two datasets were obtained by the Harlan J. Smith Telescope (HJST) with \add{the} Robert G. Tull Coud\'e Spectrograph \citep[TS23;][]{1995PASP..107..251T} at McDonald Observatory on 12 November 2008 UT and 4 October 2014 UT. \add{These} datasets \add{were also} used in our previous study \cite{2020PASJ...72...19W}. To extract each line profile from each spectrum, we adopted least-squares deconvolution \citep[LSD;][]{1997MNRAS.291..658D}; the observed spectrum \add{was} regarded as a convolution of a line profile and a series of delta functions. Under this method, we obtained a list of \add{the} absorption lines from the Vienna Atomic Line Database \citep[VALD;][]{2000BaltA...9..590K} to create a series of delta functions. \add{We then} derived each line profile and \add{the} error bars with the matrix calculations in \cite{2010A&A...524A...5K}.

\add{We then} have analysed three spectroscopic datasets obtained on 28 September 2016 UT, 12 January 2018 UT\add{,} and 2 January 2019 UT \add{from} the high-resolution HARPS-N spectrograph \citep[][]{2012SPIE.8446E..1VC}\add{,} which is mounted at the Telescopio Nazionale Galileo (TNG). These three datasets have been extracted and published in \cite{2021A&A...653A.104B}, \add{which were} used \add{in} the extracted line profile series \add{in this study}.

We also We then included two aditional spectral datasets. One was \add{an} extracted dataset taken by HJST/TS23 on 11 December 2016 UT. \add{Data from} 2016 also include\add{d} 10 in-transit spectra. The other dataset was obtained from the 188 cm telescope with HIgh Dispersion Echelle Spectrograph \citep[HIDES;][]{1999PYunO....S..77I} at Okayama Astro-Complex (OAC) in Japan on 27 December, 2019 UT. We utilized a wavelength range from $4980$\AA\ to $6220$\AA\, except for the Na D lines and \add{wavelength} regions around bad pixels. We reduced these spectral data by \add{the} subtracting bias and dark features, flat dividing, and performing one-dimensional spectrum and wavelength calibration. We then utilized their continua\add{,} as well as the \add{HDS} process. We selected alp Leo\add{, a} rapidly-rotating star \citep[$V \sin i_s \sim 300$ km s$^{-1}$;][]{2002ApJ...573..359A}\add{,} to erase the Earth's atmospheric absorption lines. \add{We then} shifted the spectra to the barycentric frame using \texttt{PyRAF} \citep{2012ascl.soft07011S}. Finally, we extracted each line profile of each exposure using the LSD.

\begin{table*}
\caption{Details \add{on the spectral datasets}}
\label{result_SP}
\centering
\small
\begin{tabular}{l l c c c c c}
\hline
  Date (UT) & Instrument & Number of spectra & Exposure time (s)& Resolution &SNR at 5500\AA&Reference\\
      \hline
      12 Nov 2008& HJST/TS23 & 13 & 900 &60,000& 140$^{\dag}$&\cite{2010MNRAS.407..507C}\\
      19 Oct 2011& Subaru/HDS & 35 & 600 (33 spectra), 480 (2 spectra) &110,000& 160$^{\dag}$&\cite{2020PASJ...72...19W}\\
      4 Oct 2014& HJST/TS23 & 21 & 900 &60,000& 280$^{\dag}$&\cite{2015ApJ...810L..23J}\\
      28 Sep 2016 & TNG/HARPS-N & 40 & 600 &115,000& 110$^{\ddag}$&\add{\cite{2021A&A...653A.104B}} \\
      11 Dec 2016 & HJST/TS23 & 21 & 900 &60,000& 250$^{\dag}$&This work \\
      12 Jan 2018 & TNG/HARPS-N & 23 & 900 &115,000& 170$^{\ddag}$&\add{\cite{2021A&A...653A.104B}} \\
      2 Jan 2019 & TNG/HARPS-N & 33 & 600 &115,000& 120$^{\ddag}$&\add{\cite{2021A&A...653A.104B}} \\
      27 Dec 2019& OAC/MuSCAT & 12 & 1200 &65,000& 50$^{\dag}$&This work\\
\hline
\hline
\multicolumn{2}{l}{$^{\dag}$ SNR per pixel}\\
\multicolumn{2}{l}{$^{\ddag}$ SNR per extracted pixel}\\
\end{tabular}
\end{table*}

\subsection{Extracting planetary shadow} \label{PlSh}
\add{We derived a median line profile from all exposure data including in-transit for each epoch. We applied a median line profile, not a mean profile, because it avoided the effects of outliers.}
We subtracted the median line profile from each exposure line profile to calculate the time\add{-}series of the line profile residuals. \add{Both the} planetary shadow due to the transit \add{of WASP-33b} and a striped pattern due to the non-radial pulsations on the surface of WASP-33 \citep{2010MNRAS.407..507C} \add{were present in the line profile residuals}. \citet{2011A&A...526L..10H} showed that the pulsation period was approximately 68 min \add{based on} photometric observation\add{s}. However, determin\add{ing} the period from the Doppler tomographic results \add{was difficult owing to} the irregular patterns.

\begin{figure*}
\centering
  \includegraphics[width=155mm]{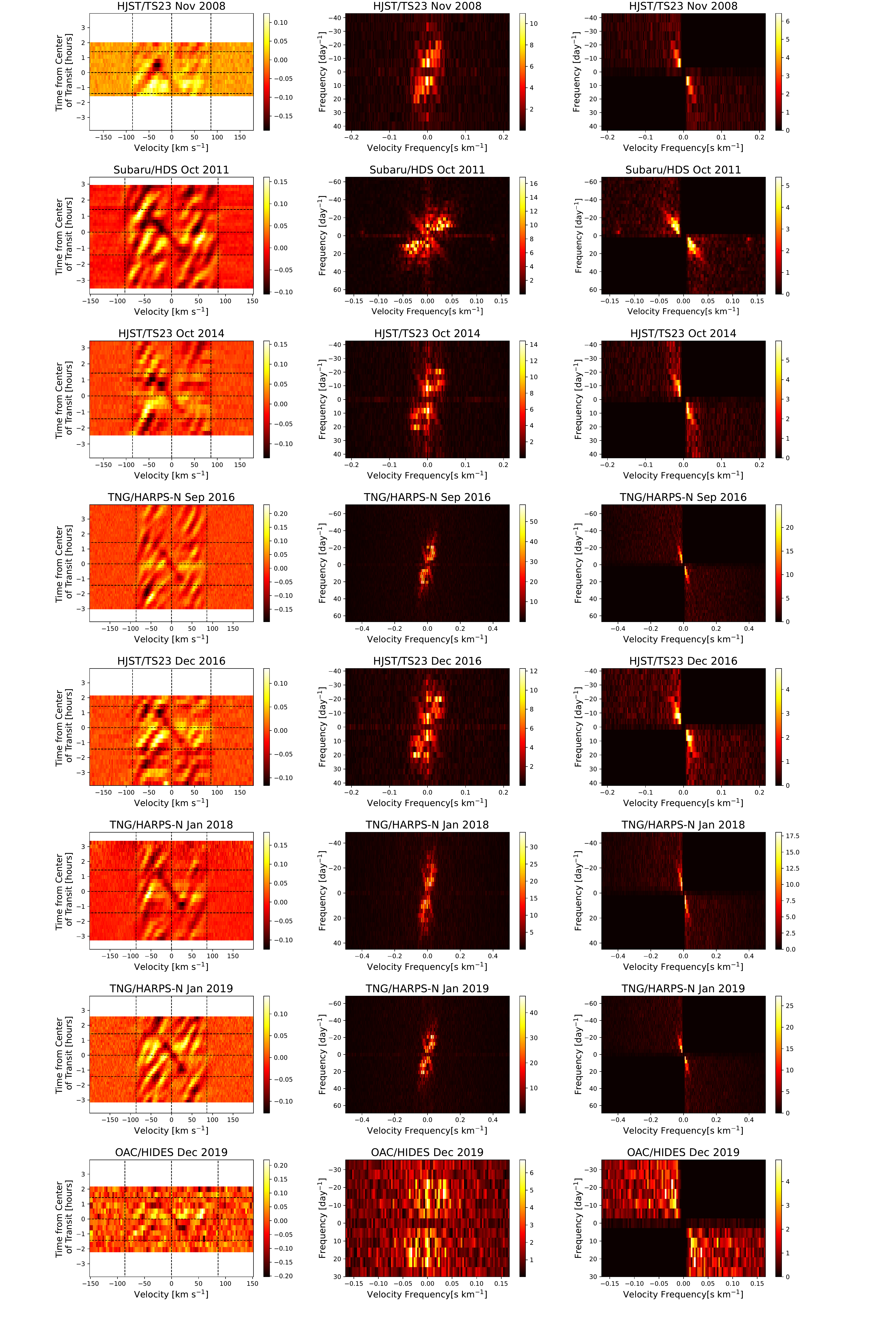}
\vspace*{-0.8cm} 
\caption{Doppler tomographic datasets and Fourier filters. Left column: observed residuals \add{for the} line profile series. The vertical dotted lines show $v=0$ and $\pm  v \sin i_{s}$. The bottom, middle, and upper horizontal dotted lines show the beginning, middle, and end of the WASP-33b's transit, respectively. Middle column: Fourier spaces after Fourier transform of the residuals \add{for} the line profile series. These colour scales are shown as square roots. The faint narrow structure from the bottom right to the upper left is a component of WASP-33b's planetary transit. \add{In contrast, t}he bright-wide structure from the bottom left to the upper right is a pulsation component. Right column: filtered Fourier space \add{such} that only the transit component remains.}
\label{figbff}
\end{figure*}

To extract only the planetary shadow, we applied a Fourier filtering technique \citep{2015ApJ...810L..23J} because the planet is retrograde, \add{while} the pulsations are prograde. First, we performed a two-dimensional Fourier transform. There were components derived from the pulsations in the Fourier space in the first and third quadrants, and \add{the} planetary shadow's components in the second and fourth quadrants. Second, we created a filter in which we set unity in \add{the} two diagonal quadrants, including power from the planetary shadow, and zero in the other quadrants, including power from the pulsation with a Hann function between these quadrants. Finally, we multiplied the Fourier space by the filter and performed an inverse Fourier transform on the filtered Fourier space to extract the planetary shadow. Figure \ref{figbff} \add{illustrates these procedures}.

\subsection{Deriving parameters} \label{DB}
To obtain the best-fit values and uncertainties of \add{the} transit parameters, we adopted the Markov chain Monte Carlo (MCMC) method using the code \texttt{EMCEE} \citep{2013PASP..125..306F}.

We modelled a planetary shadow \add{via} a convolution between a rotational broadening profile and a Gaussian line profile owing to intrinsic broadening, thermal broadening, and micro-turbulence. \add{The detailed} equations to derive the model of the planetary shadow are described in the appendix of \cite{2020PASJ...72...19W}\add{.}  \add{We then} applied the same filter to the model of the planetary shadow following the procedures described in Section \ref{PlSh}.

We fitted the observed residuals of the five datasets to the models with 21 parameters using MCMC: \add{the} $\lambda$, $b$ and transit mid-time $T_{c}$ of each epoch, $V\sin i_{s}$, $R_{p}/R_{s}$, $a/R_{s}$, two quadratic limb darkening coefficients, and the FWHM of the Gaussian line profile. The limb darkening coefficients \add{were} derived \add{via} the triangular sampling method \add{reported in} \citet{2013MNRAS.435.2152K}, \add{with} $q_{1}$, and $q_{2}$. We estimated that $q_{1}$ and $q_{2}$ of HDS, TS23, and HIDES \add{were} equivalent. They \add{were} calculated from the stellar parameters, \add{i.e.}, the effective temperature $T_{\mathrm{eff}}$, surface gravity log $g$, and metallicity. We set the priors of $\lambda$ and $b$ for all epochs and the FWHM as uniform functions \add{while that of} the priors of the other parameters \add{were set} as Gaussian priors. For \add{the} values and widths of \add{the} Gaussian priors, we set the priors \add{for} $R_{p}/R_{s}$ and $a/R_{s}$ based on the values and uncertainties \add{reported in} \citet{2013AA...553A..44K}\add{;} the priors of each $T_{c}$ of each epoch from $P_{\mathrm{orb}}$ in \citet{2014A&A...561A..48V} and $T_{0}$ in \citet{2019A&A...622A..71V}\add{;} \add{those} of $q_{1}$, and $q_{2}$ calculated \add{via} $\mathrm{PyLDTk}$ \citep{Parviainen2015, Husser2013}\add{;} and \add{those} of $V\sin i_{s}$ from \citet{2015ApJ...810L..23J}.

For the fitting, we maximised the logarithm of the posterior probability\add{,} $\ln L_{\mathrm{post}}$\add{:}
\begin{eqnarray}
\label{eq1}
\ln L_{\mathrm{post}} = -\sum_{i} \frac{(O_{i}-C_{i})^{2}}{\sigma^{2}_{i}} - \sum_{j} \frac{(p_{j}-\mu_{j})^{2}}{s^{2}_{j}},
\end{eqnarray}
where $O_{i}$ is the data, $C_{i}$ is the model, $\sigma_{i}$ is the error for the $i$th data point, $p$ is the parameter value at the gained iteration of the Markov chain, $\mu$ is the value from the literature, and $s$ is the uncertainty from the literature. \add{Index} $j$ denote\add{s} the parameters of the Gaussian priors. 
We set the range of the uniform prior of each $\lambda$ to $-180 \mathrm{deg} <\lambda<-90 \mathrm{deg}$ and that of each $b$ as $-1<b<1$.
To converge these parameter values, we ran 4,000 steps, cut off the first 2,000 steps as burn-in, and iterated this set 100 times. \add{Figures \ref{figMCMC} and \ref{figMCMC_2} in Appendix \ref{MCMC_res} plot the} posterior distributions.

\section{Photometric Measurement} \label{sec:PH}
\subsection{Photometric Observations of WASP-33b's Transit}
We observed the transit of WASP-33b by photometry using two instruments \add{with} multicolour simultaneous cameras: Multicolor Simultaneous Camera for studying Atmospheres of Transiting exoplanets \citep[MuSCAT;][]{2015JATIS...1d5001N} on the 188 cm telescope at OAC and MuSCAT2 \citep{2019JATIS...5a5001N} on \add{the} Telescopio Carlos S\'{a}nchez (TCS) 1.52 m telescope at the Teide Observatory (OT). MuSCAT has three channels for \add{the} $g^{'}_{2}$ (400-550 nm), $r^{'}_{2}$ (550-700 nm), and $z_{\mathrm{s},2}$ (820-920 nm) bands. In contract, MuSCAT2 contains four channels for \add{the} $g^{'}_{2}$, $r^{'}_{2}$, $z_{\mathrm{s},2}$, and $i^{'}_{2}$ (700-820 nm) bands. These bands are the Astrodon Photometrics Generation 2 -type Sloan filters. We obtained the dataset \add{from} MuSCAT on 5 November 2017 UT. We set the exposure times of \add{the} $g^{'}_{2}$, $r^{'}_{2}$ and $z_{\mathrm{s},2}$ bands for 4, 4, and 10 s, respectively. We also obtained the dataset \add{from} MuSCAT2 on 11 October 2018 UT. The exposure time of \add{the} $g^{'}_{2}$, $r^{'}_{2}$, $i^{'}_{2}$, and $z_{\mathrm{s},2}$ \add{was} 3, 2, 5, and 12 s, respectively.

To produce the light curves of WASP-33b, aperture photometry was performed using the pipeline proposed by \citet{2011PASJ...63..287F}. This process determines the stellar barycentre in every frame. We used BD+36 488, the second brightest star image in the frame, as a companion star to correct the atmospheric extinction. Next, the pipeline calculated the shift in the stellar position relative to the reference frame. \add{This} approach photometers the target star and a comparison star with a fixed aperture radius. Here\add{,} we set the aperture radii to 36 and 40 pixels for MuSCAT and MuSCAT2, respectively. After the sky background in the torus area centred on the stellar barycentr, the pipeline \add{subtracted} the sky background from WASP-33b's flux and \add{that of the comparison}. Finally, the WASP-33b light curve was obtained by dividing its flux by the comparison star flux.

\subsection{Light Curve Fitting} \label{LCF}
To measure WASP-33b's impact parameter $b$ in 2017 and 2018, we constructed light curve models with a Gaussian process using the Python code \texttt{exoplanet} \citep{exoplanet:exoplanet}. WASP-33b's light curve not only \add{showed} dimming by the transit, but also a short sinusoidal-wave-like feature due to the stellar pulsations. Thus, following a previous study by \citet{2015ApJ...810L..23J}, we applied a Matern 3/2 kernel $\bf{K}_{ker}$, whose element is expressed as \add{follows:}
\begin{eqnarray}
\label{eq2}
k_{i,j}= \alpha^2\left(1+\frac{\sqrt{3}|t_i-t_j|}{l}\right)\exp \left(-\frac{\sqrt{3}|t_i-t_j|}{l}\right)+\sigma_{i} ^{2}\delta_{i,j}
\end{eqnarray}
for the Gaussian process. $i$ and $j$ denote the orders of the photometric observation's data, $t_i$ and $t_j$ are the observation times, $\alpha$ and $l$ are the hyper parameters indicating the amplitude and timescale of the stellar variations, respectively, and $\sigma$ is the error of data point $i$. 

We then fitted the light curves on two epochs to the models with the following 30 parameters using MCMC: baseline $B$ for each light curve, $b$ and $T_{c}$ of each epoch, two quadratic limb darkening coefficients $u_1$, $u_2$, and $\alpha$ of each band, $R_{p}/R_{s}$, $P_{\mathrm{orb}}$, $a/R_{s}$, and $l$. We set the priors of $B$ and $b$ for both epochs, and $R_{p}/R_{s}$ as uniform functions. \add{We then set} the priors of  $T_{c}$ for each epoch, $u_1$, $u_2$, and $\alpha$ for each band, $P_{\mathrm{orb}}$, $a/R_{s}$, and $l$ \add{as} Gaussian priors. For the values and widths of \add{the} Gaussian priors, we referred to the values and widths of Gaussian priors from \citet{2015ApJ...810L..23J} for $\alpha$ and $l$ and \citet{2014A&A...561A..48V} for $P_{\mathrm{orb}}$; the others \add{were obtained} in the same manner as the spectral analysis. We set the logarithm of the likelihood $\ln P_{\mathrm{like}}$ as
\begin{eqnarray}
\label{LH_GP}
\ln P_{\mathrm{like}} = -\frac{1}{2}(\ln |\bf{K}_{ker}|+\bf{r} ^{T}\bf{K}_{ker} ^{-1}\bf{r})
\end{eqnarray}
because we adopted Gaussian process for this fitting \citep{10.5555/1162254}.
\add{Here,} $\bf{r}$ is a series of residuals obtained by subtracting the model data from the observation data. In the MCMC process \add{in} \texttt{PyMC} \citep{exoplanet:pymc3}, we ran 1,000 steps, cut off the first 500 steps as burn-in, and iterated this set 20 times. \add{Figures \ref{figMCMC_Ph} and \ref{figMCMC_Ph_2} in Appendix \ref{MCMC_res} plot the} posterior distributions.

\section{Fitting with Nodal Precession Model} \label{FitMod}
The angular momentum of WASP-33b's planetary orbit $|\overrightarrow{L_{p}}|$ (= $2 \pi M_{p} a^{2}/P_{\mathrm{orb}}$) is \add{significantly} smaller than the stellar rotational angular momentum of its host star $|\overrightarrow{L_{s}}|$; $|\overrightarrow{L_{p}}|/|\overrightarrow{L_{s}}|$ is $\sim 0.05$ using \add{the values} of $|\overrightarrow{L_{s}}|$ from \citet{2011Ap&SS.331..485I}, \add{those} of $M_{p}$ from \citet{2015AA...578L...4L}, $a$, and $P$ from \citet{2010MNRAS.407..507C}. In this case, we can regard the stellar rotational axis as a stable vector and calculate the changes in $b$ and $\lambda$
\begin{eqnarray}
b(t)&=&\frac{a}{R_s} \left(\cos \psi \cos i_s + \sin \psi \sin i_s \cos \theta(t) \right)
\label{impact_pro} \\
\tan \lambda(t)&=&\frac{\sin \psi \sin \theta(t)}{\sin \psi \cos i_s \cos \theta(t)-\cos \psi \sin i_s}.
\label{lambda_pro}
\end{eqnarray}
Here $\theta$, the nodal angle, can be expressed as \add{follows:}
\begin{eqnarray}
\theta (t)&=&-\frac{3 \pi J_{2} R_s^{2} \cos \psi}{P_{\mathrm{orb}}a^2}t +\theta_0,
\label{theta_pro}
\end{eqnarray}
where the slope of Equation \ref{theta_pro} is the precession speed from \citet{2013ApJ...774...53B}. 

\add{We then} fit\add{ted} the model \add{in} Equations \ref{impact_pro} and \ref{lambda_pro} with the values measured by \add{the} MCMC using \texttt{PyMC}. We considered $\psi$, $\theta (t=2008)$, $i_s$, and $J_{2}$ as free parameters and set their priors as uniform functions. Here, we \add{set} $\theta$ in 2008 as $\theta_0$, \add{i.e.,} the initial value of $\theta$ in Equation \ref{theta_pro}. For this fitting, we set the logarithm of the likelihood $\ln P_{\mathrm{like}}$ as:
\begin{eqnarray}
\ln L_{\mathrm{like,pre}}=-\frac{1}{2}\sum_{i} \frac{(O_{\mathrm{mes}, i}-C_{\mathrm{mod},i})^{2}}{\sigma_{\mathrm{mes},i}^{2}}
\label{chi_pre}
\end{eqnarray}
where $O_{\mathrm{mes}, i}$ is the measured value of $\lambda$ and $b$ of each epoch, $C_{\mathrm{mod},i}$ is the model value of $\lambda$ and $b$, and $\sigma_{\mathrm{mes},i}$ is the uncertainty of the measured $\lambda$ and $b$, respectively. We consider that the values $b$ in 2017 and 2018 were zero, as shown in Figure \ref{figMCMC_Ph}.
We ran 20,000 steps, cut off the first 10,000 steps as burn-in, and iterated this set 20 times.
The posteriors from the MCMC are shown in Figure \ref{result_system}, and the values are listed in Table \ref{result_2}. \add{Figure \ref{change}} exhibit\add{s} the changes in $\lambda$ and $b$ \add{for} WASP-33b.

\section{Results} \label{result}
\begin{figure*}
\centering
  \includegraphics[width=155mm]{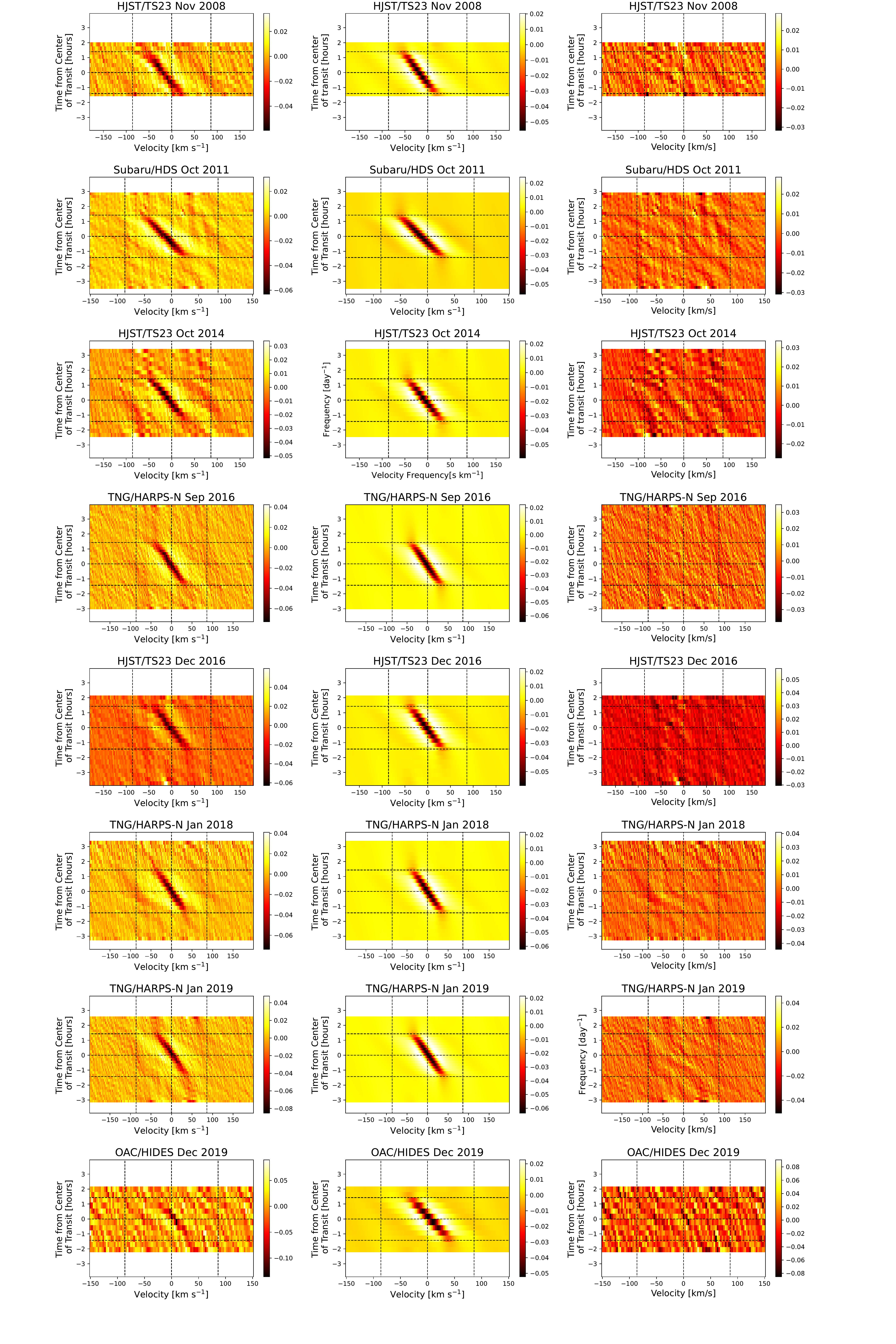}
  \vspace*{-0.8cm} 
\caption{Fitting \add{of the} filtered residual data by MCMC. \add{Same} type of colour scale as \add{that in} the left column \add{of} Figure \ref{figbff}. Left column: \add{r}esidual data remain\add{ing} as a planetary shadow. Middle column: \add{f}iltered models of a planetary shadow using \add{the} best-fit values. Right column: the difference between the first row and the second rows.}
\label{resulr_ph}
\end{figure*}

\add{Figure \ref{resulr_ph} shows the} line profile residuals and the best-fitted filtered models. \add{Table \ref{result_1} lists the} best values of $\lambda$ and $b$. Our results for $\lambda$ and $b$ in 2014 \add{were} in excellent agreement with the values of \citet{2015ApJ...810L..23J} within 1$\sigma$, whereas those in 2008 were marginally consistent with \citet{2015ApJ...810L..23J} within $2\sigma$. Moreover, our results for $\lambda$ in 2016 and 2018 from HARPS-N \add{were} consistent with those of \cite{2021A&A...653A.104B}, whereas our $\lambda$ in 2019 from HARPS-N differs from that in \cite{2021A&A...653A.104B} by $\sim 2\sigma$.

\begin{figure*}
\centering
  \includegraphics[width=1.\textwidth]{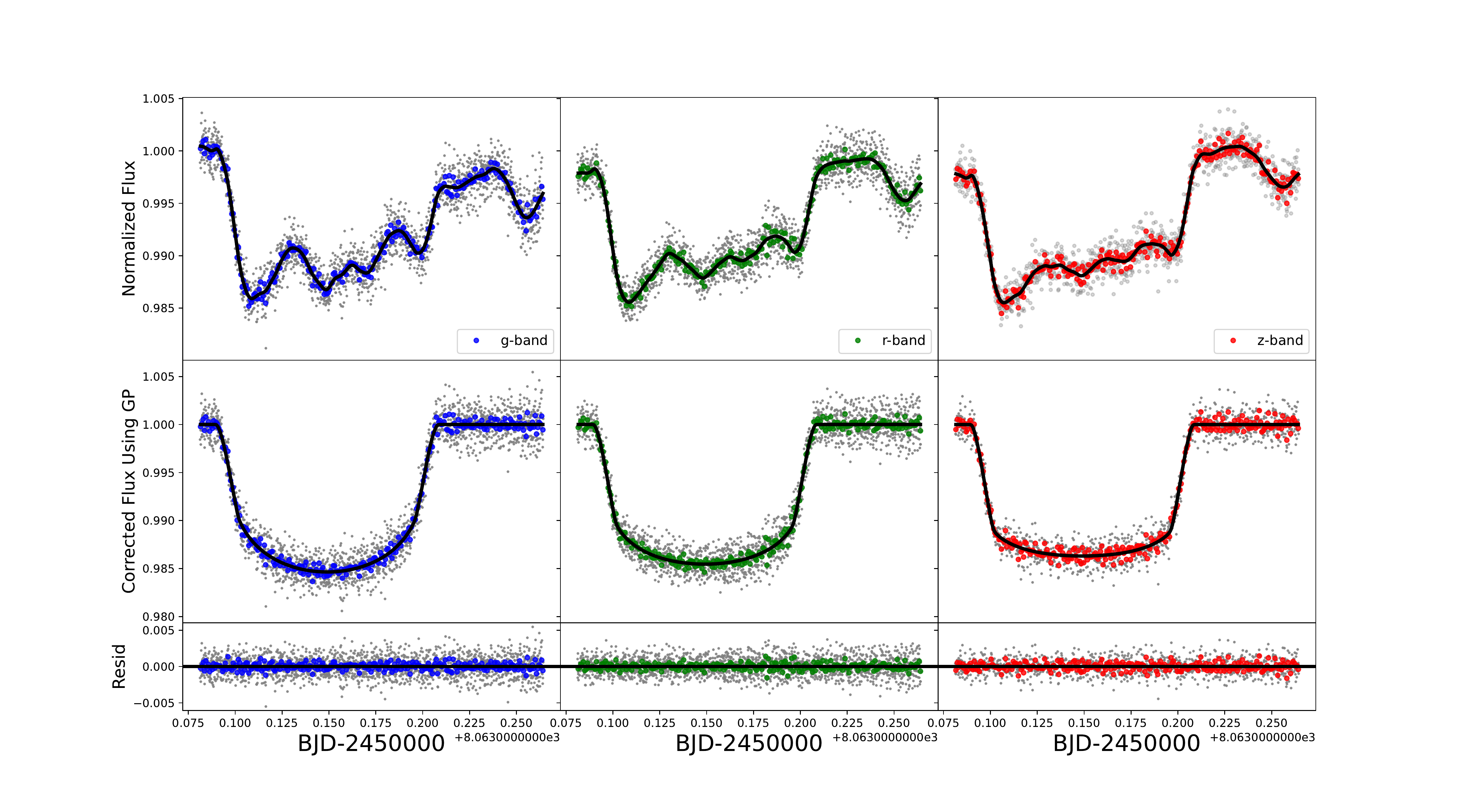}
\caption{Light curves of MuSCAT in 2017 with the original cadences (gray points) and 1-minute bins (coloured points). Top row: light curves from \add{the} photometric observation data. The black solid lines represent the models from the MCMC fitting. Middle row: light curves subtracted using \add{the} Gaussian process. Bottom row: residuals between the observed data and model data.}
\label{resulr_LC_1}
\end{figure*}

\begin{figure*}
\centering
  \includegraphics[width=1.\textwidth]{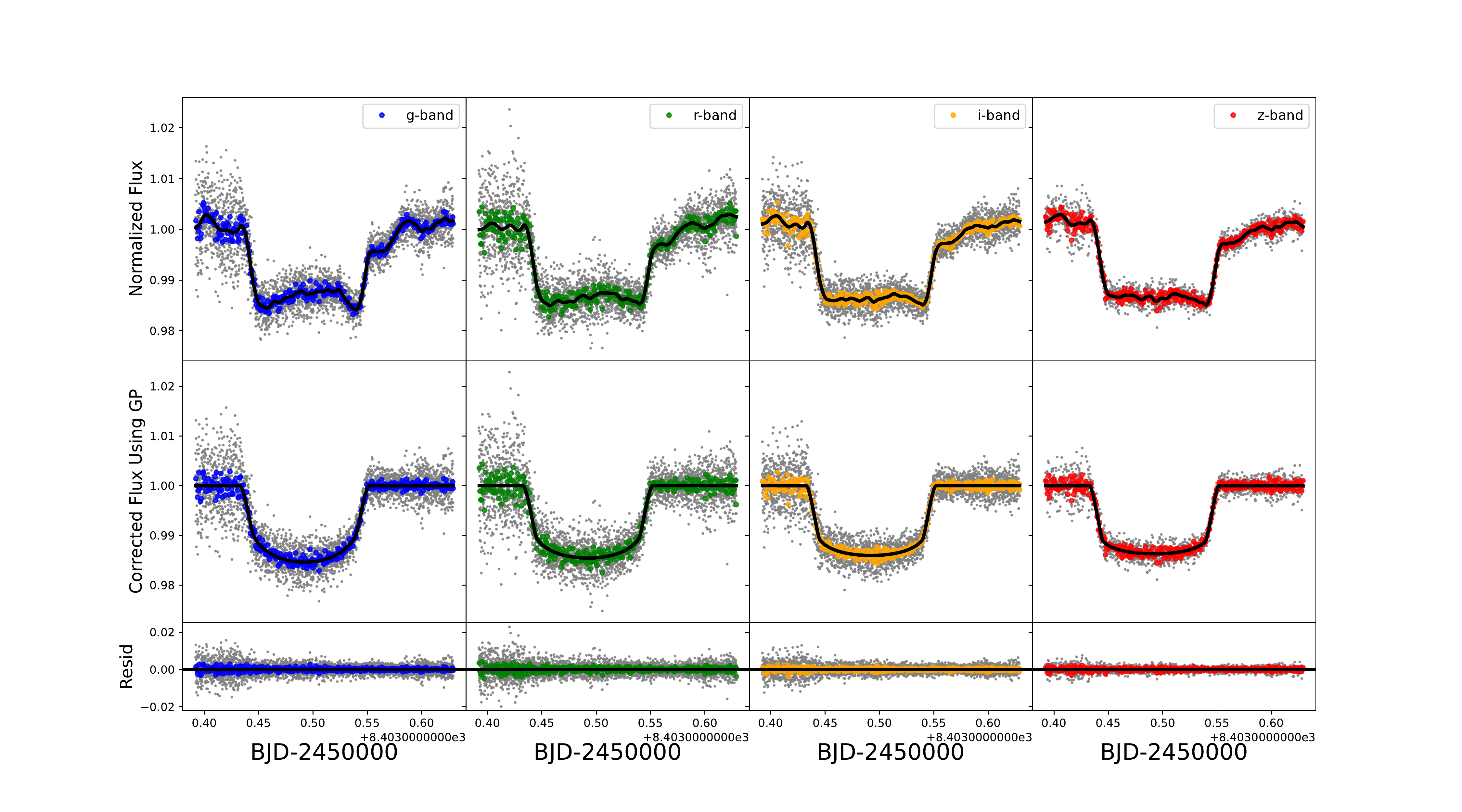}
\caption{Similar light curves as Figure \ref{resulr_LC_1}, but for \add{the} MuSCAT2 dataset.}
\label{resulr_LC_2}
\end{figure*}

\add{Figures \ref{resulr_LC_1} and \ref{resulr_LC_2} show the} best-fit light curve models. \add{Table \ref{result_1} lists the} ranges of $b$ from MuSCAT and MuSCAT2. The posteriors of both $b$ in Figure \ref{figMCMC_Ph} exhibit a truncated normal distribution with a minimum value of $\sim$ 0. Hence, we set 1$\sigma$ as a 68\% confidence interval \add{based on} the minimum value \add{of} 0 because \add{they were identical,} but \add{with} opposite signs of impact parameters\add{, which yielded} the same transit light curves.

\renewcommand{\arraystretch}{1.1}
\begin{table*}
\caption{Measured \add{p}arameters \add{for} WASP-33b}
\label{result_1}
\centering
\small
\begin{tabular}{l c c c c c}
\hline
Date (UT)& $\lambda$ (deg)& $b$& $T_c (\mathrm{BJD_{TDB}})$&Method & Instrument\\
      \hline
      12 Nov 2008  & $-111.30^{+0.76}_{-0.77}$ & $0.2398^{+0.0062}_{-0.0058}$ &$2454782.92502\pm 0.00016$& Doppler tomography & HJST/TS23 \rule[-1mm]{0mm}{5mm}\\
      19 Oct 2011& $-113.96\pm 0.30$ & $0.1578\pm 0.0027$&$2455853.96863^{+0.00014}_{-0.00015}$& Doppler tomography& Subaru/HDS \rule[-1mm]{0mm}{5mm}\\
      4 Oct 2014& $-113.00\pm 0.37$ & $0.0845\pm+0.0031$ &$2456934.77139\pm0.00015$& Doppler tomography & HJST/TS23 \rule[-1mm]{0mm}{5mm}\\
      28 Sep 2016& $-111.39\pm0.23$ & $0.0413\pm 0.0019$ &$2457660.59250\pm0.00014$& Doppler tomography & TNG/HARPS-N \rule[-1mm]{0mm}{5mm}\\
      11 Dec 2016& $-111.32^{+0.49}_{-0.47}$ & $0.0432\pm 0.0039$ &$2457733.78452\pm0.00015$& Doppler tomography & HJST/TS23 \rule[-1mm]{0mm}{5mm}\\
      5 Nov 2017& - & $|b|<0.132$ &$2458063.14903\pm0.00015$& Photometry &OAC/MuSCAT\rule[-1mm]{0mm}{5mm}\\
      11 Oct 2018& - & $|b|<0.067$ &$2458403.49234\pm0.00014$& Photometry &TCS/MuSCAT2\rule[-1mm]{0mm}{5mm}\\
      12 Jan 2018& $-111.46\pm 0.28$ & $0.0034^{+0.0024}_{-0.0023}$ &$2458131.46135\pm0.00015$& Doppler tomography & TNG/HARPS-N \rule[-1mm]{0mm}{5mm}\\
      2 Nov 2019& $-111.64\pm 0.28$ & $-0.0272^{+0.0020}_{-0.0021}$ &$2458486.44296\pm0.00016$& Doppler tomography & TNG/HARPS-N \rule[-1mm]{0mm}{5mm}\\
      27 Dec 2019 & $-112.24^{+0.97}_{-1.02}$ & $-0.0592^{+0.0066}_{-0.0065}$ &$2458845.08379^{+0.00015}_{-0.00016}$& Doppler tomography & OAC/HIDES \rule[-1mm]{0mm}{5mm}\\
\hline
\hline
\end{tabular}
\end{table*}

\begin{figure}
\centering
  \includegraphics[width=85mm]{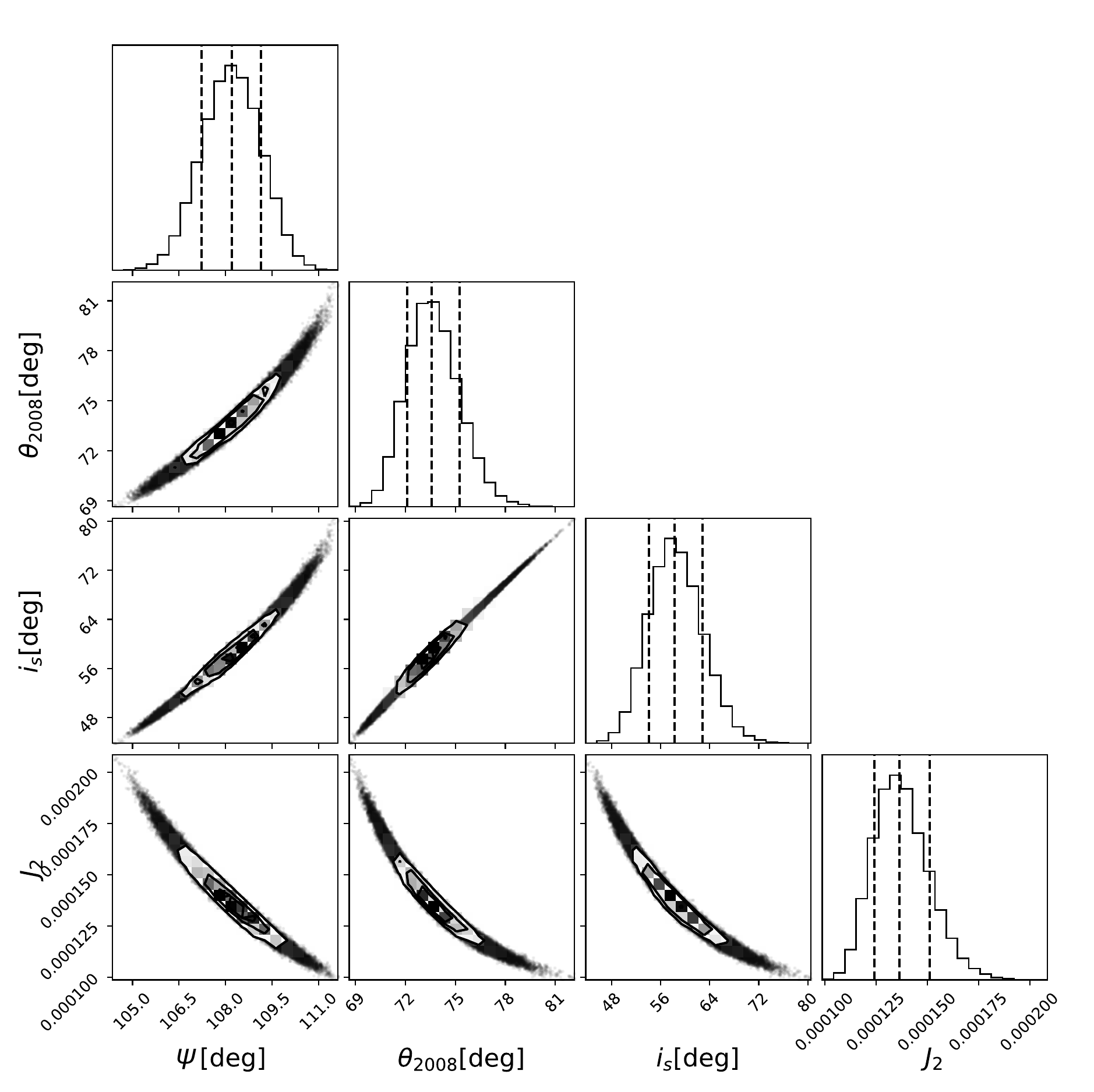}
\caption{MCMC corner plots for \add{the} architecture angles, $\psi$, $\theta_{\mathrm{2008}}$ and $i_s$, and stellar quadrupole moment $J_2$ of \add{the} WASP-33b System}
\label{result_system}
\end{figure}

\begin{figure*}
\centering
 \begin{minipage}{0.45\hsize}
   \includegraphics[width=85mm]{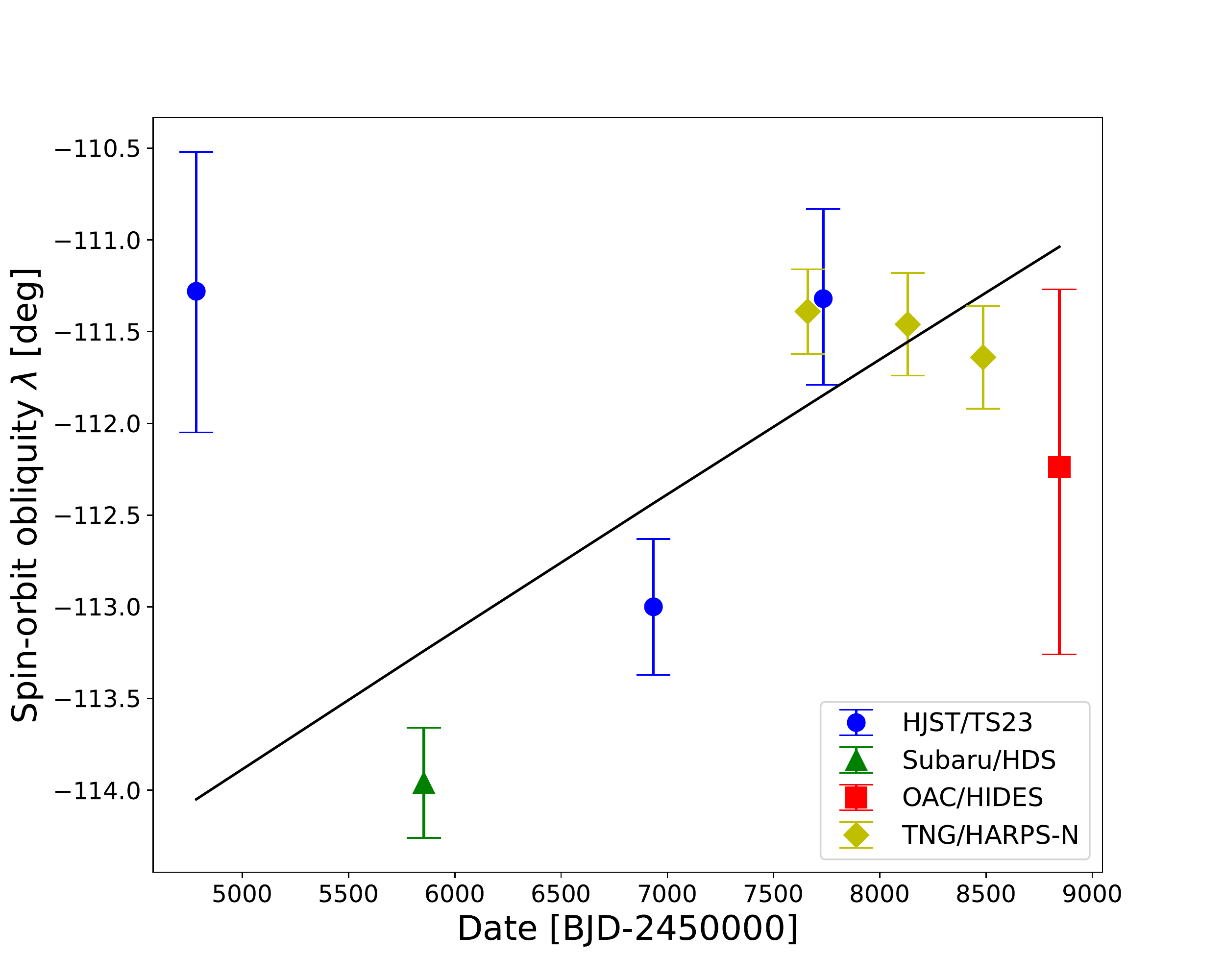}
 \end{minipage}
 \begin{minipage}{0.45\hsize}
   \includegraphics[width=85mm]{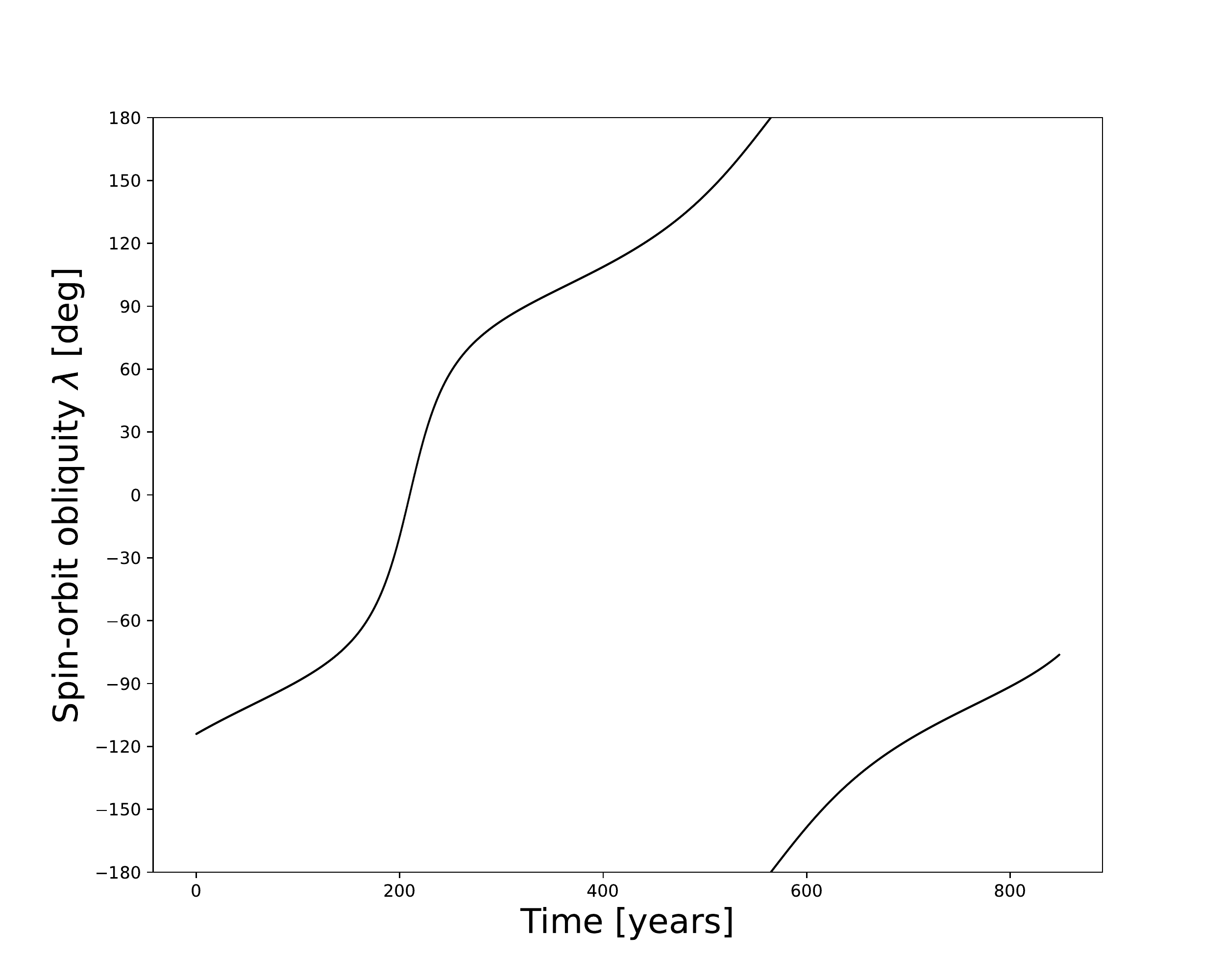}
 \end{minipage}\\
 \begin{minipage}{0.45\hsize}
   \includegraphics[width=85mm]{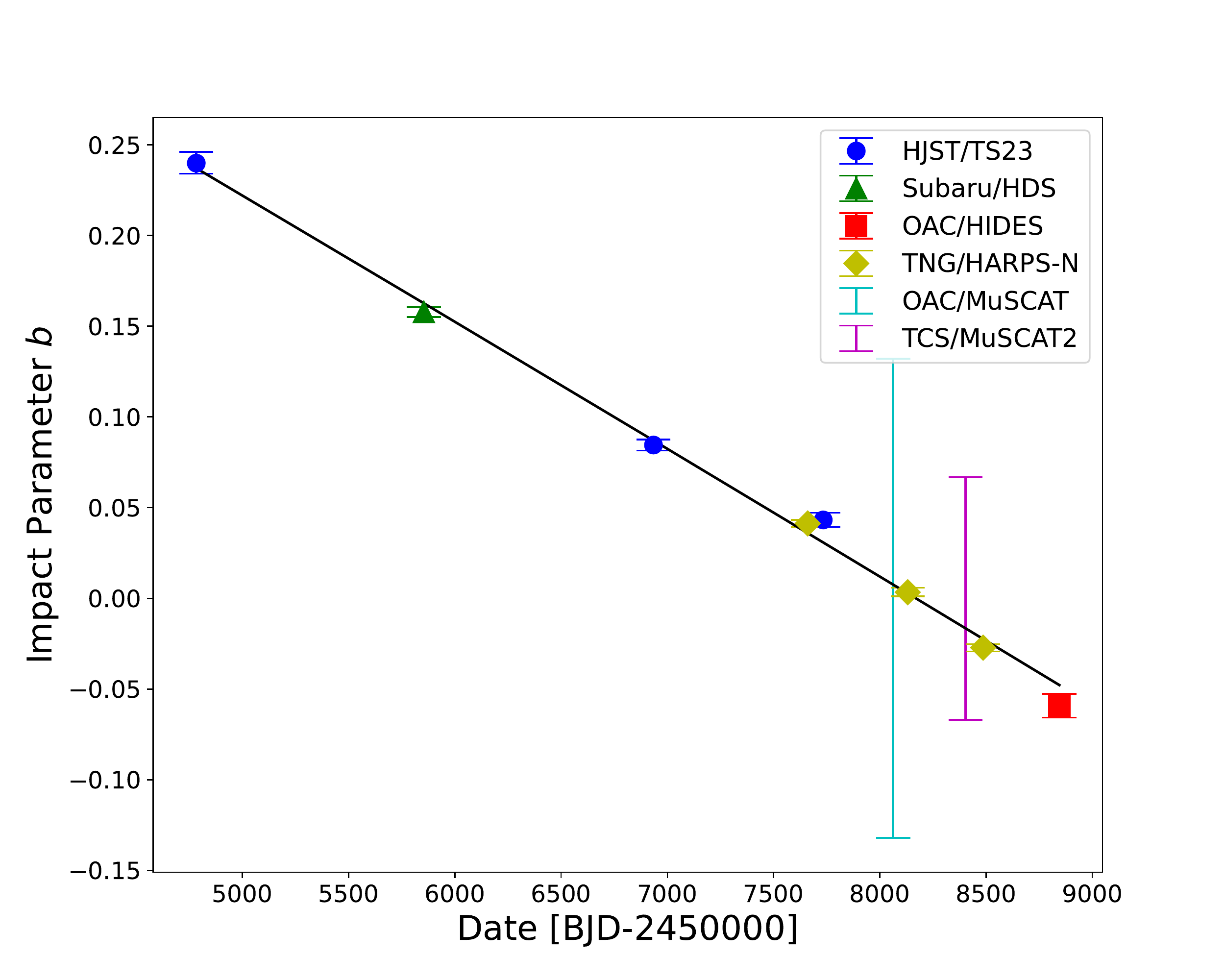}
 \end{minipage}
 \begin{minipage}{0.45\hsize}
   \includegraphics[width=85mm]{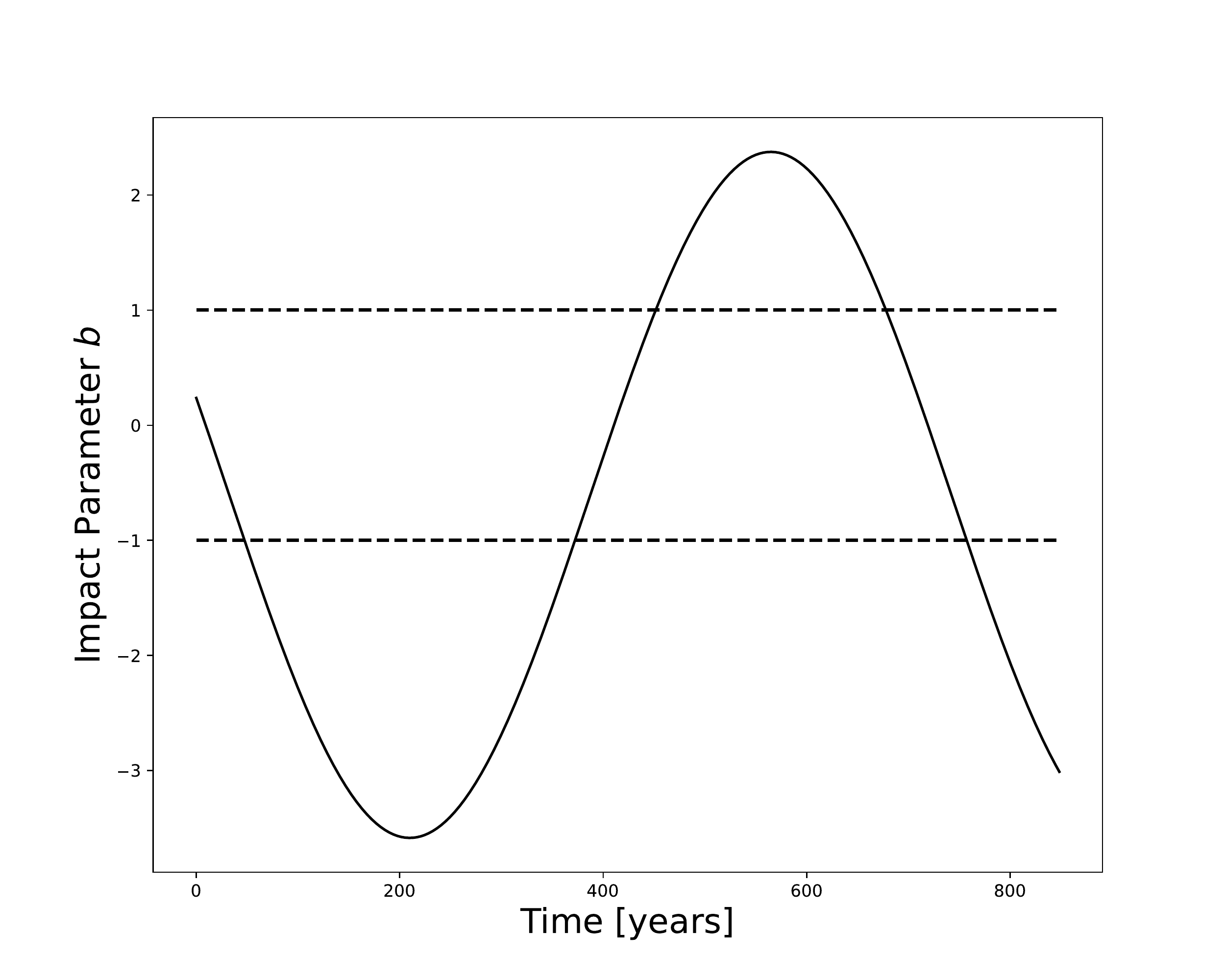}
 \end{minipage}
 \caption{Changes \add{in} $\lambda$ (upper row) and $b$ (lower row). The left and right columns show the the short\add{-} and long\add{-}term \add{changes}, respectively. The x-axis of the right column is the time in years from the epoch of 2008. The blue circles show values from the HJST/TS23 \add{datasets}, the green triangles are values from Subaru/HDS, the red squares are values from OAC/HIDES, the cyan line is the value range from OAC/MuSCAT, the magenta line is \add{the} value range from TCS/MuSCAT2, and the black solid lines represent the model. In the bottom-right figure, \add{the} two black\add{-}dashed lines show the edges of the stellar disk of WASP-33b.}
 \label{change}
\end{figure*}

Moreover, \add{Figure \ref{result_system} presents} posteriors from MCMC with the nodal precession model\add{.} \add{Table \ref{result_2}lists} the values of $\psi$, $\theta (t=2008)$, $i_s$, and $J_{2}$ . Here, we note that \citet{2016MNRAS.455..207I} may have consider\add{ed} $\psi$ as a variable value. \add{Figure \ref{change} shows the} change in $\lambda$ and $b$ \add{for} WASP-33b.

\begin{table*}
\caption{Calculated \add{p}arameters \add{for} WASP-33b}
\label{result_2}
\centering
\small
\begin{tabular}{l c c c c}
\hline
  Date& $\psi$ (deg)&$J_2$& $i_s$ (deg)& $\theta_{2008}$ (deg)\\
      \hline
      Results of this study&$108.19^{+0.95}_{-0.97}$ &$(1.36^{+0.15}_{-0.12})\times 10^{-4}$&$58.3^{+4.6}_{-4.2}$&$73.6^{+1.7}_{-1.5}$ \rule[-2mm]{0mm}{5mm}\\
      \citet{2016MNRAS.455..207I} &
      $99^{+5}_{-4}$ (in 2008), $103^{+5}_{-4}$ (in 2014)
      & $(2.1^{+0.8}_{-0.5})\times 10^{-4}$ &$142^{+10}_{-11}$ & -  \rule[-2mm]{0mm}{5mm}\\
      \citet{2020PASJ...72...19W}& - &$(9.14\pm 0.51)\times 10^{-5}$& $96^{+10}_{-14}$ & -  \rule[-2mm]{0mm}{5mm}\\
      \add{\citet{2021A&A...653A.104B}} & $113.99\pm 0.22$ &$(6.73\pm 0.22)\times 10^{-5}$& $90.11\pm 0.12$ & - \rule[-2mm]{0mm}{5mm}\\
      \citet{2021arXiv210903250D}& $108.3^{+19.0}_{-15.4}$ & - & $69.8^{+4.0}_{-3.2}$ & - \rule[-2mm]{0mm}{5mm}\\
\hline
\hline
\end{tabular}
\end{table*}

\section{Discussion} \label{Discuss}
We have inspected the nodal precession of WASP-33b with more observations than \add{that used} in previous studies \citep{2015ApJ...810L..23J, 2020PASJ...72...19W, 2021A&A...653A.104B}. This is the first study to verify the nodal precession \add{based on} both Doppler tomographic observatioadd{s} and transit photometry. The errors from the transit photometric observations \add{were} large. However, with the change in the impact parameter (see the left bottom part \add{in} Figure \ref{change}), the results from the transit photometry \add{were} consistent with the predicted values from the decreasing trend \add{for} the Doppler tomographic observations. This indicates that transit photometry can \add{simultaneously} contribute to the measurement\add{s} of the nodal precession using Doppler tomographic data.
In Figure \ref{change}, although the impact parameter of WASP-33b \add{appeared} to change linearly, the change in $\lambda$ may not be along the model of the nodal precession. Thus, \add{we should} observe WASP-33b \add{via} Doppler tomography to clarify whether its $\lambda$ increase\add{s} \add{based on} the model or decrease\add{s} from 2021.

\subsection{Comparison of Stellar Spin Inclination and Quadrupole Moment with Previous Studies}
Our $i_s$ value disagrees with those in previous studies \citep{2015ApJ...810L..23J, 2016MNRAS.455..207I, 2020PASJ...72...19W, 2021A&A...653A.104B}\add{, as} derived from the nodal precession by $\sim$3 $\sigma$ or more. However, our value is \add{similar} to that of \citet{2021arXiv210903250D}, \add{despite values that} are $\sim$2 $\sigma$ \add{different}. Notably, \cite{2021arXiv210903250D} derived WASP-33b's $i_s$ from light curve TESS photometric data considering its oblateness and gravity darkening.

The derived stellar quadrupole moment \add{of WASP-33b was} $J_2 = (1.36^{+0.15}_{-0.12}) \times 10^{-4}$. This value agrees with that of \citet{2016MNRAS.455..207I} within 1.5 $\sigma$, which is larger than those of the other previous studies (add{$>$} 3 $\sigma$) and smaller than the theoretical value ($J_2 = 3.8\times 10^{-4}$) calculated by \cite{2011Ap&SS.331..485I}.

\add{One of the possible reasons for disagreements with the values of $i_s$ and $J_2$ could be the difference in the nodal precession model. \cite{2016MNRAS.455..207I} and \cite{2020PASJ...72...19W} used time variation models for other orbital parameters: the ascending node $\Omega$ and the orbital inclination to the apparent equatorial plane $I$, illustrated in Figure 4 of \cite{2020PASJ...72...19W}, and calculated from $\lambda$ and $b$. Then, \cite{2021A&A...653A.104B} estimated the change in the inclination angle $i_p (=\arccos{(bR_s/a)})$ as a linear function. In this study, we directly used the accurate time variation models of $\lambda$ and $b$, such that our derived values for $i_s$ and $J_2$ would be more accurate than those in previous studies. However, we should clarify the cause of the short-term variation of $\lambda$ to create more detailed nodal precession model.}

\add{In contrast, there is a probability that adding datasets also causes the disagreements. In this study, we found a short-term variation in $\lambda$, which may have been decreasing since 2016, although the reason remains unclear. Therefore,} we \add{have} to obtain more datasets of WASP-33b's transits to \add{more accurately} determine their values.

\subsection{Orbital Evolution of WASP-33b}

We found the real spin-orbit obliquity of WASP-33b $\psi=108.19^{+0.95}_{-0.97}$ deg. This uncertainty is larger than that reported by \citet{2021A&A...653A.104B} \add{owing to} the difference in the precession model, which shows that $d\lambda / dt$ of WASP-33b is always positive, while that of \citet{2021A&A...653A.104B} allows $d\lambda / dt = 0$ at a certain time. The derived value indicates the possibility that WASP-33b has \add{experienced} \add{the planet-planet scattering \citep{2008ApJ...686..580C} or Kozai migration \citep{2007ApJ...669.1298F}, which are the mechanisms that cause the misaligned orbit.} The existence of the WASP-33b's companion star is necessary \add{for clarify} distinguish\add{ing} these two evolution\add{ary} models. Nevertheless, Doppler tomographic observations and transit photometry cannot detect the companion stars.

\citet{2016ApJ...827....8N} found a companion \add{candidate} \add{for} WASP-33\add{,} estimated \add{as} a dwarf star or brown dwarf \add{and} located at 238 AU ($P_{\mathrm{orb}} \sim 3,300$ yrs) from the host star \add{based on} direct imaging. However, \add{this candidate} has not yet been confirmed \add{owing to negligible} proper motion. Thus, additional direct imaging observation\add{s} \add{are} required to confirm whether the companion moves in the same proper motion as the host star or not. If the stellar companion candidate\add{,} with 0.1 $M_{\star}$\add{,} revolves in a circular orbit ($e=0$) and WASP-33b was formed near the snow line, $\sim 5$ AU ($P_{\mathrm{orb}} \sim 12$ yrs), the planet \add{experienced} a Kozai oscillation with the Kozai cycle period\add{,} $P_{\mathrm{Kozai}}$\add{,} \add{of} 13 Myrs \add{based on the} following \add{expression:}
\begin{equation}
\label{KCP}
P_{\mathrm{Kozai}}=\frac{M_s P_c^2}{M_c P_b}(1-e_c^2)^{3/2},
\end{equation}
where $M_s$, $M_c$, $P_c$, $P_b$, and $e_c$ are the stellar mass, companion's mass, companion's period, planetary mass, and companion's eccentricity, respectively \citep{2007ApJ...670..820W}. Although we have to consider that the distance from the host star to the companion star is the sky-projected distance and the eccentricity of the companion star may exist, the Kozai cycle period \add{could} be shorter than the age of WASP-33 ($\sim$ 100 Myrs). Therefore, the stellar companion may have caused the Kozai mechanism \add{for} WASP-33b.

\subsection{Nodal Precession Speed}
We calculated the nodal precession speed $\dot{\theta}=0.507^{+0.025}_{-0.022}$ deg year$^{-1}$ and its period $P_{\mathrm{pre}}=709^{+33}_{-34}$ years, \add{where} the precession \add{was} faster than that of \citet{2016MNRAS.455..207I} ($\dot{\theta}=0.37^{+0.04}_{-0.03}$). \add{we then found} that WASP-33b transits in front of the host star for only $\sim$ 20 \% of the \add{entire} nodal precession period, which indicates that it is rare to discover WASP-33b as a transiting planet. This implies that WASP-33b began transiting in $1977\pm 2$ and will stop transiting in $2055\pm 2$.

\subsection{Nodal Precession Observations of Other Hot Jupiters around Hot Stars}
Doppler tomography has confirmed 17 hot Jupiters around hot stars \add{to date}. The real spin-orbit obliquities of seven of these planets (WASP-33b, Kepler-13Ab, KELT-9b, KELT-17b, MASCARA-1b, MASCARA-4b, and WASP-189b) have been identified thus far, \add{which are} nearly vertical (60 deg < $\psi$ < 120 deg). \add{The remaining hot Jupiters} reveal \add{that} only their projected spin-orbit obliquities were obtained. Three hot Jupiters around hot stars with $\lambda \sim 90$ deg, \add{i.e.,} KELT-26b \citep[$\lambda=91.3^{+6.5}_{-6.3}$ deg;][]{2020AJ....160..111R}, HAT-P-70b \citep[$\lambda=113.1^{+5.1}_{-3.4}$ deg;][]{2019AJ....158..141Z}\add{,} and TOI-1518b \citep[$\lambda=-119.66^{+0.98}_{-0.93}$ deg;][]{2021AJ....162..218C}, are valuable for detecting their observable nodal precessions to derive their $\psi$. Even if $\lambda$ is near 0 deg or 180 deg, we can observe the change in \add{the} transit trajectory and measure $\psi$ \add{near} 90 deg when the star rotates nearly pole-on for the line of sight. However, when $\lambda$ is near 0 or 180 deg and the star rotation axis is almost perpendicular to the line of sight, the transit trajectory \add{barely} moves because $\psi$ should also be near 0 deg or 180 deg. In this case, we can estimate $\psi$ as $\lambda$. Therefore, we should \add{regularly} observe the nodal precessions of these planets around their hot stars to measure their $\psi$ using Doppler tomography and transit photometry.

Kepler-13Ab, orbiting around an A-type star, is the another hot Jupiter whose nodal precession has been detected by only transit photometries \citep{2011ApJ...736L...4S, 2011ApJS..197...10B}. \add{P}revious studies \add{have} measured $\psi$ of this hot Jupiter \add{via} gravity-darkened transit photometry, but the values were different between these two results ($\psi=60\pm 2$ deg in \cite{2015ApJ...805...28M} and $\psi=29\pm 1$ deg in \cite{2018AJ....155...13H}). \add{Al}though Kepler-13Ab is likely to have evolved with Kozai migration \add{owing} to its companion star, Kepler-13B \citep{2012A&A...544L..12S}, the value of its $\psi$ should be verified by adding Doppler tomographic observations and transit photometries to obtain an accurate histogram of the derived $\psi$. \add{As} \citet{2014ApJ...790...30J} measured \add{the} $\lambda$ \add{of Kepler-13Ab} in 2014 \add{via} Doppler tomography, an additional transit spectroscopic observation enabled us to detect the change in $\lambda$ and then to derive its $\psi$ independently from the gravity-darkened transit photometry.

\cite{2020ApJ...888...63A}, \cite{2020A&A...643A..94L}, \cite{2020AJ....160....4A}, and \cite{2021arXiv210905031H} measured \add{the} $\psi$ of MASCARA-4b ($\psi=104^{+7}_{-13}$ deg), WASP-189b ($\psi=85.4\pm 4.3$ deg), KELT-9b ($\psi=87^{+10}_{-11}$deg), and MASCARA-1b ($\psi=72.1^{+2.5}_{-2.4}$ deg), respectively, using gravity-darkened transit photometry. 
\cite{2016AJ....152..136Z} derived the $\psi$ of KLET-17b ($\psi=116 \pm 4$ deg) by the technique using \add{the} differential rotation of a host star.

These planets are hot Jupiters around A-type stars\add{:} their nodal precessions are yet to be detected. Observing the nodal precessions of these hot Jupiters is important \add{for verifying} the values of their $\psi$. This observation can also contribute to the investigation\add{s} \add{on} the internal structure of hot stars by deriving the values of their $J_2$.

Although there are 17 hot Jupiters around hot stars whose projected spin-orbit obliquities have been measured, \add{this} number \add{remains} too small to statistically determine the orbital evolution tendency. \citet{2021ApJ...916L...1A} found that planets around solar-like stars with large projected spin-orbit obliquities are likely to revolve on polar orbits ($\psi \sim 90$ deg); however, the tendency remains unclear \add{for} hot Jupiters around hot stars because the real spin-orbit obliquities of only six of them have been revealed.
If normal planet-disk interaction is the main migration, the distribution should gather at $\psi = 0$ deg \citep{2011MNRAS.412.2790L}.
If the orbital evolution by planet-planet scattering is the majority, the orbits are likely to incline \add{at approximately} $\psi = 60$ deg \citep{2011ApJ...742...72N}.
When Kozai migration is the primary evolution, $\psi$ \add{has} a wide range from 10 deg to 140 deg \citep{2015ApJ...799...27P}.
\add{As} \citet{2018ApJS..239....2B} predicted that the TESS mission could find 500 hot Jupiters around A-type stars from the 2-year observation, \add{increasing} the number of hot Jupiters around hot stars \add{will become possible} by validating TESS planet candidates.
Further observations \add{could} lead to uncovering the \add{detailed} $\psi$ distribution around hot stars.

\section{Conclusion} \label{Concl}
We analysed \add{the} nodal precession \add{of WASP-33b via} Doppler tomography and transit photometry using various high-dispersion spectrographs including Subaru/HDS, HJST/TS23, OAC/HIDES, and TNG/HARPS-N, and two multicolour simultaneous cameras, OAC/MuSCAT and TCS/MuSCAT2.
Based on the observed change in the projected spin-orbit obliquity $\lambda$ and the impact parameter $b$, we modelled the nodal precession of WASP-33b and derived the real spin-orbit obliquity of WASP-33b as $\psi = 108.19^{+0.95}_{-0.97}$ deg. 
Compared \add{with} the results of previous studies, the results of the near-polar orbit did not change. However, our value \add{for} $\psi$ differ\add{ed} from that of \cite{2021A&A...653A.104B} by $>3 \sigma$. \add{This discrepancy may be caused by whether or not $psi$ is a constant value. We assumed $psi$ as a constant while \cite{2021A&A...653A.104B} did not; \cite{2021A&A...653A.104B} adopted $psi$ from its value at the 2011 epoch.} 
We also simultaneously derived the stellar spin inclination and the stellar gravitational quadrupole moment of WASP-33 as $i_s = 58.3^{+4.6}_{-4.2}$ deg, and $J_2=(1.36^{+0.15}_{-0.12}) \times 10^{-4}$, respectively. These results \add{differed} by $> 3 \sigma$ from those of previous studies \add{on} nodal precession, except for \add{the} $J_2$ \add{value reported in} \cite{2016MNRAS.455..207I} ($\sim 2.1\sigma$). 

\add{A likely reason for these discrepancies is the different nodal precession models. For the first time, we applied the accurate time variation models for $\lambda$ and $b$ to fit the nodal precession. Therefore, our derived values of $\psi$, $i_s$, and $J_2$ are the most accurate to date. Moreover,} additional datasets may have updated these parameter values\add{, thus causing} these differences. Thus, acquiring more datasets will allow \add{for} the derivation of more accurate values of $\psi$, $i_s$, and $J_2$.

We calculated the nodal precession speed $\dot{\theta}=0.507^{+0.025}_{-0.022}$ deg year$^{-1}$ and its period $P_{\mathrm{pre}}=709^{+33}_{-34}$ years, \add{which revealed} that WASP-33b transits in front of the host star for only $\sim$ 20 \% of the \add{entire} nodal precession period. \add{Based on} this result, we speculate that WASP-33b \add{began} transiting in 1977$\pm$2\add{; we} forecast that it \add{will} finish transiting in 2055$\pm$2.

\add{The} TESS survey \add{should} help us increase \add{our ability to count} number of hot Jupiters around hot stars in the future. Applying the proposed methodology to newly discovered hot Jupiters around hot stars is important not only \add{for characterising} each planetary system\add{,} but also \add{for discriminating} the migration mechanisms of such planets and \add{investgating} the internal structure of hot stars.

\section*{Acknowledgements}
This paper is based on data collected at the Subaru Telescope, which is located atop Maunakea and operated by the National Astronomical Observatory of Japan (NAOJ). We wish to recognize and acknowledge the very significant cultural role and reverence that the summit of Maunakea has always had within the indigenous Hawaiian community.
The paper also includes data taken at The McDonald Observatory of The University of Texas at Austin and taken at The Okayama Astrophysical Observatory.
This article is based on observations made with the MuSCAT2 instrument, developed by ABC, at Telescopio Carlos Sánchez operated on the island of Tenerife by the IAC in the Spanish Observatorio del Teide.
Pyraf is a product of the Space Telescope Science Institute, which is operated by AURA for NASA. This work has made use of the VALD database, operated at Uppsala University, the Institute of Astronomy RAS in Moscow, and the University of Vienna. We are grateful to editage (https://www.editage.jp) for English editing.
We acknowledge the GAPS Consortium (Covino et al. 2013) for providing the mean line profiles of their HARPS-N transits, and thank F. Borsa for providing spectral datasets of HARPS-N.
This work is partly supported by JSPS KAKENHI Grant Numbers JP21K20376, JP18H05439, JP20J21872, JP17H04574, JP20K14518, JST CREST Grant Number JPMJCR1761, Astrobiology Center SATELLITE Research project AB022006, and the Astrobiology Center of National Institutes of Natural Sciences (NINS) (Grant Number AB031010).
N.C.B. acknowledges funding from the European Research Council under the European Union’s Horizon 2020 research and innovation program under grant agreement no. 694513.
E. E-B. acknowledges financial support from the European Union and the State Agency of Investigation of the Spanish Ministry of Science and Innovation (MICINN) under the grant PRE2020-093107 of the Pre-Doc Program for the Training of Doctors (FPI-SO) through FSE funds.

\section*{Data Availability}

The raw data from OAC/HIDES, OAC/MuSCAT and TCS/MuSCAT2 will be shared on reasonable request to the corresponding author. 
The reduced data of Subaru/HDS were provided by N. Narita by permission, and will be shared on request to the corresponding author with permission of N. Narita.
The line profile data from HJST/TS23 were provided by M. C. Johnson by permission, and will be shared on request to the corresponding author with permission of M. C. Johnson.
The line profile data of TNG/HARPS-N were provided by F. Borsa by permission, and will be shared on request to the corresponding author with permission of F. Borsa.




\bibliographystyle{mnras}
\bibliography{example} 




\appendix

\section{MCMC results of Doppler Tomographic and Photometric Measurements} \label{MCMC_res}
\begin{figure*}
\centering
\hspace*{-2cm}
 \begin{minipage}{0.45\hsize}
   \includegraphics[width=75mm]{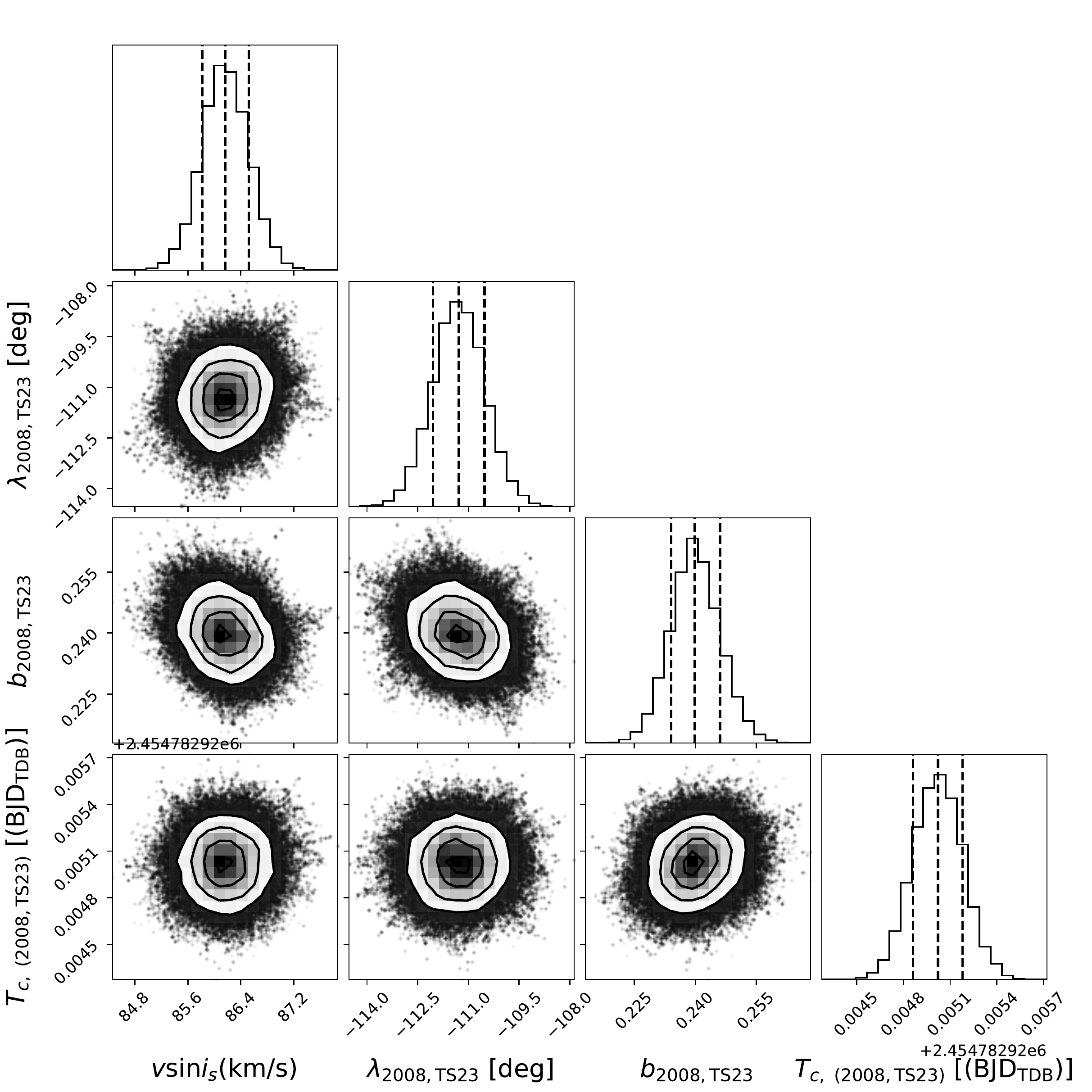}
 \end{minipage}
 \begin{minipage}{0.4\hsize}
   \includegraphics[width=75mm]{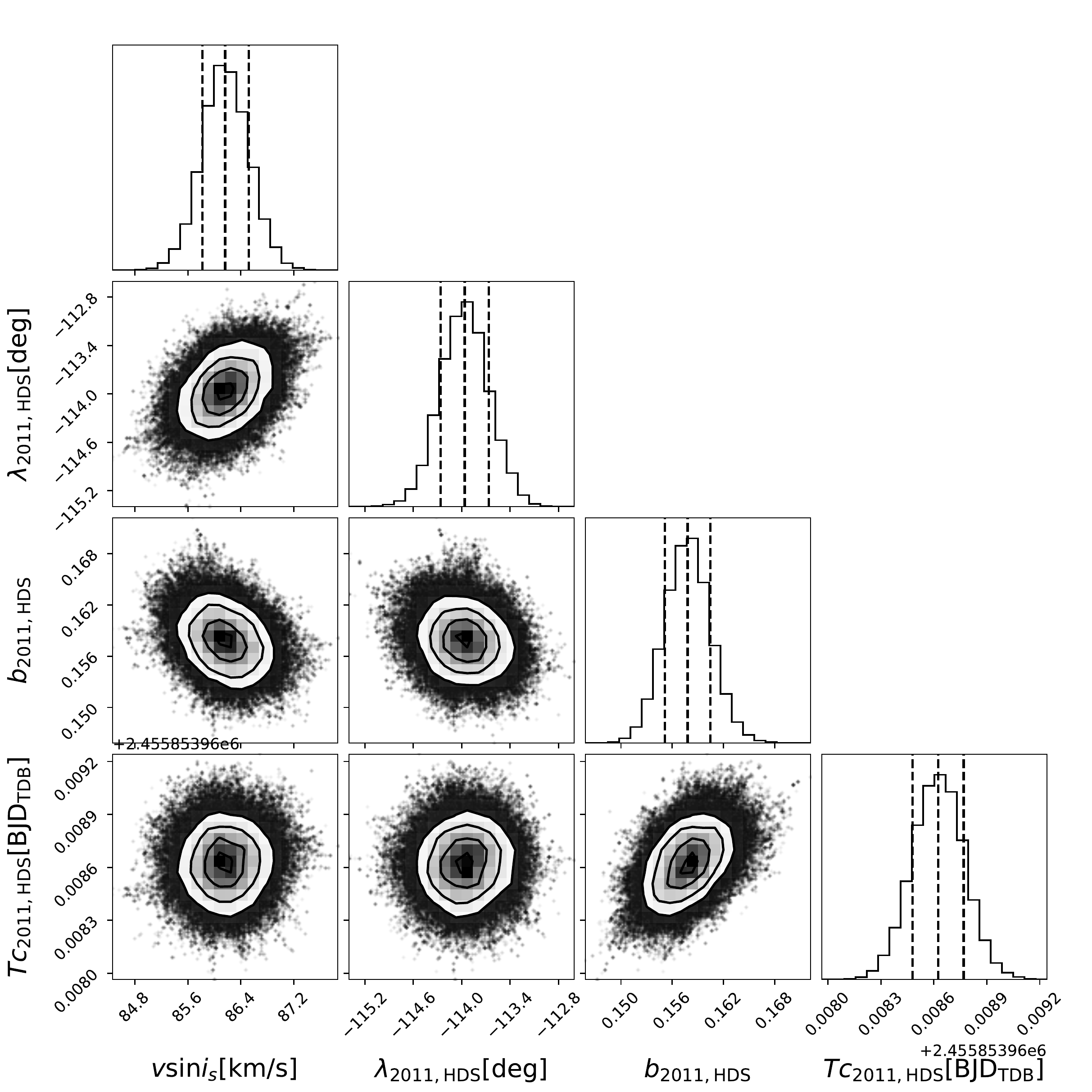}
 \end{minipage}\\
 \hspace*{-2cm}
 \begin{minipage}{0.45\hsize}
   \includegraphics[width=75mm]{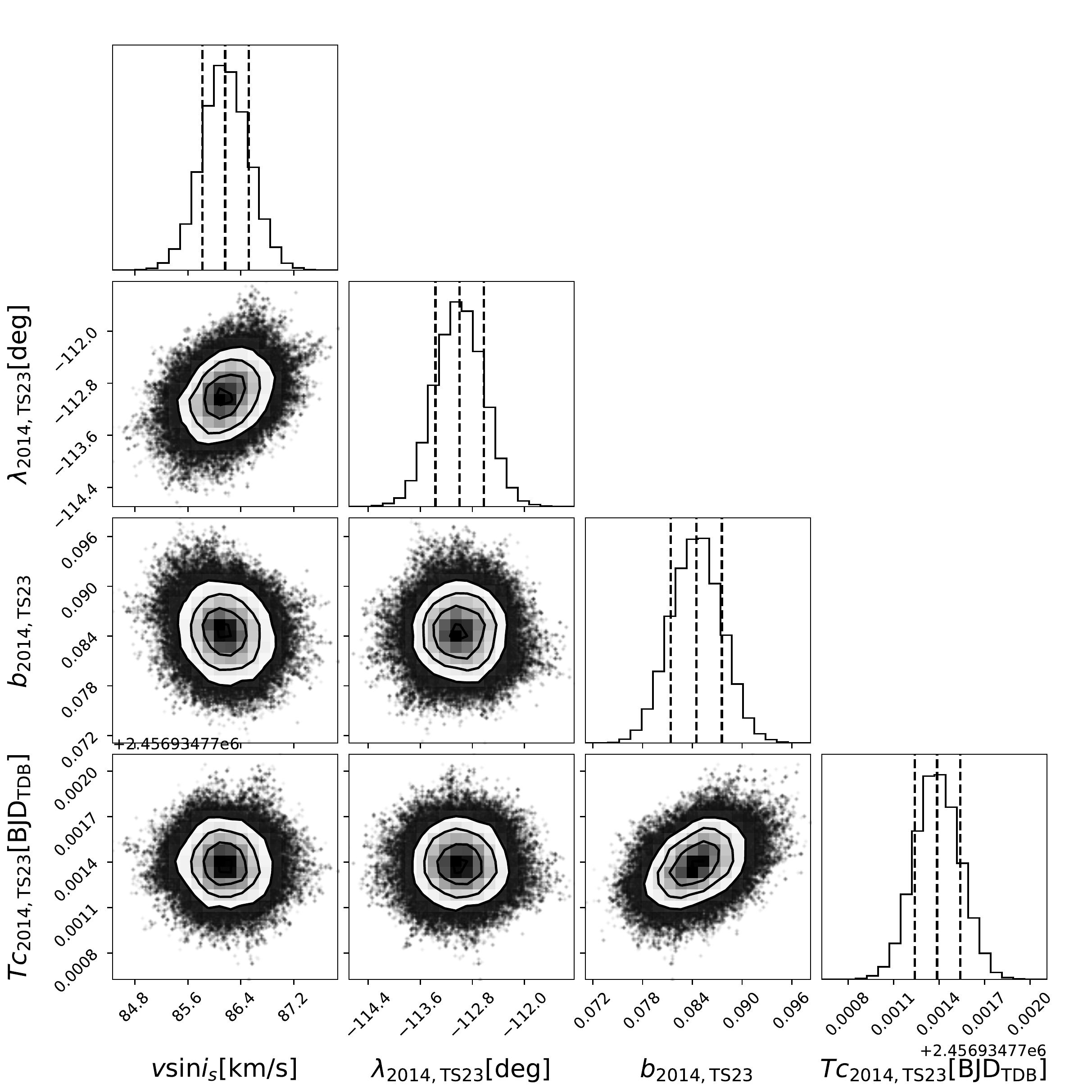}
 \end{minipage}
 \begin{minipage}{0.4\hsize}
   \includegraphics[width=75mm]{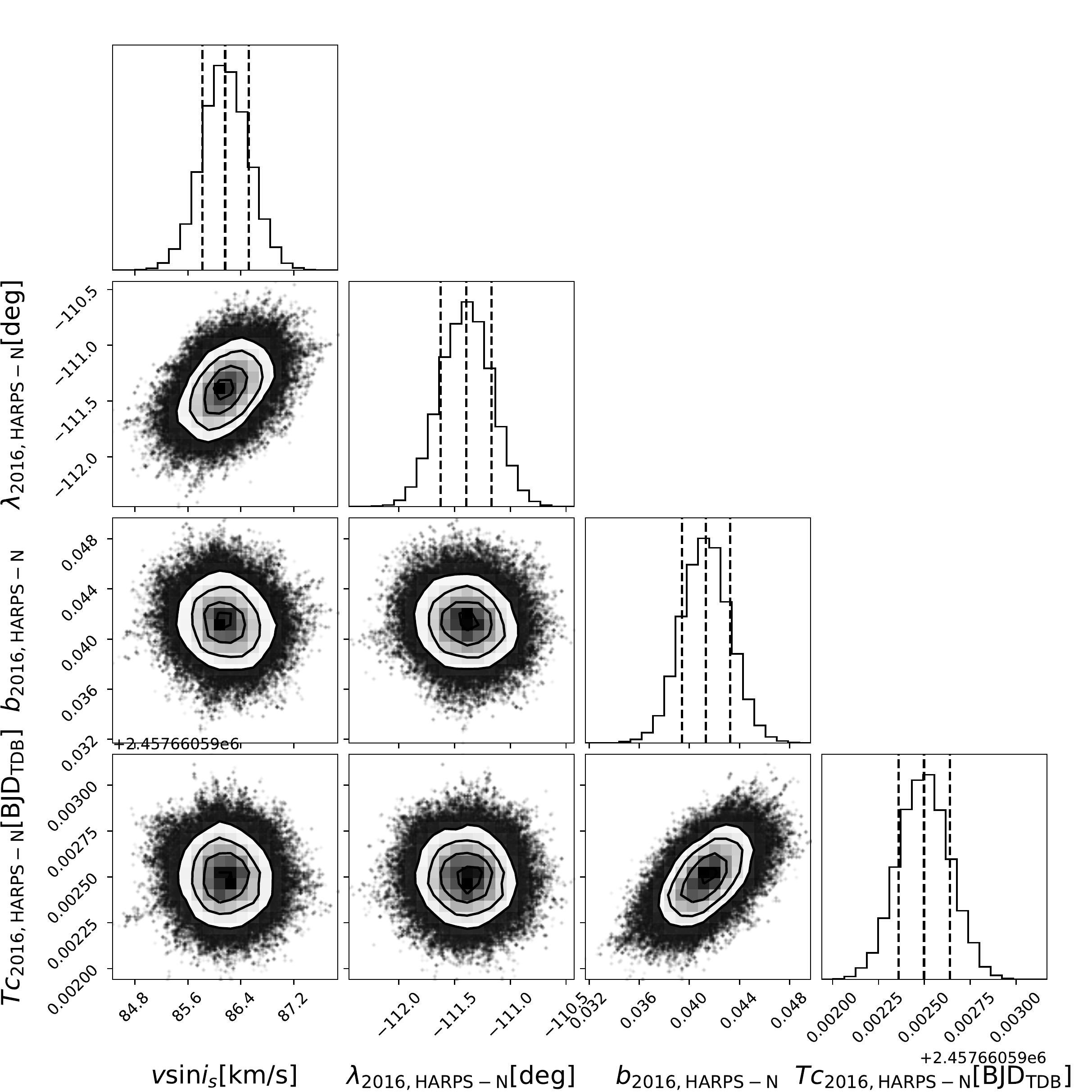}
 \end{minipage}\\
 \hspace*{-2cm}
  \begin{minipage}{0.45\hsize}
   \includegraphics[width=75mm]{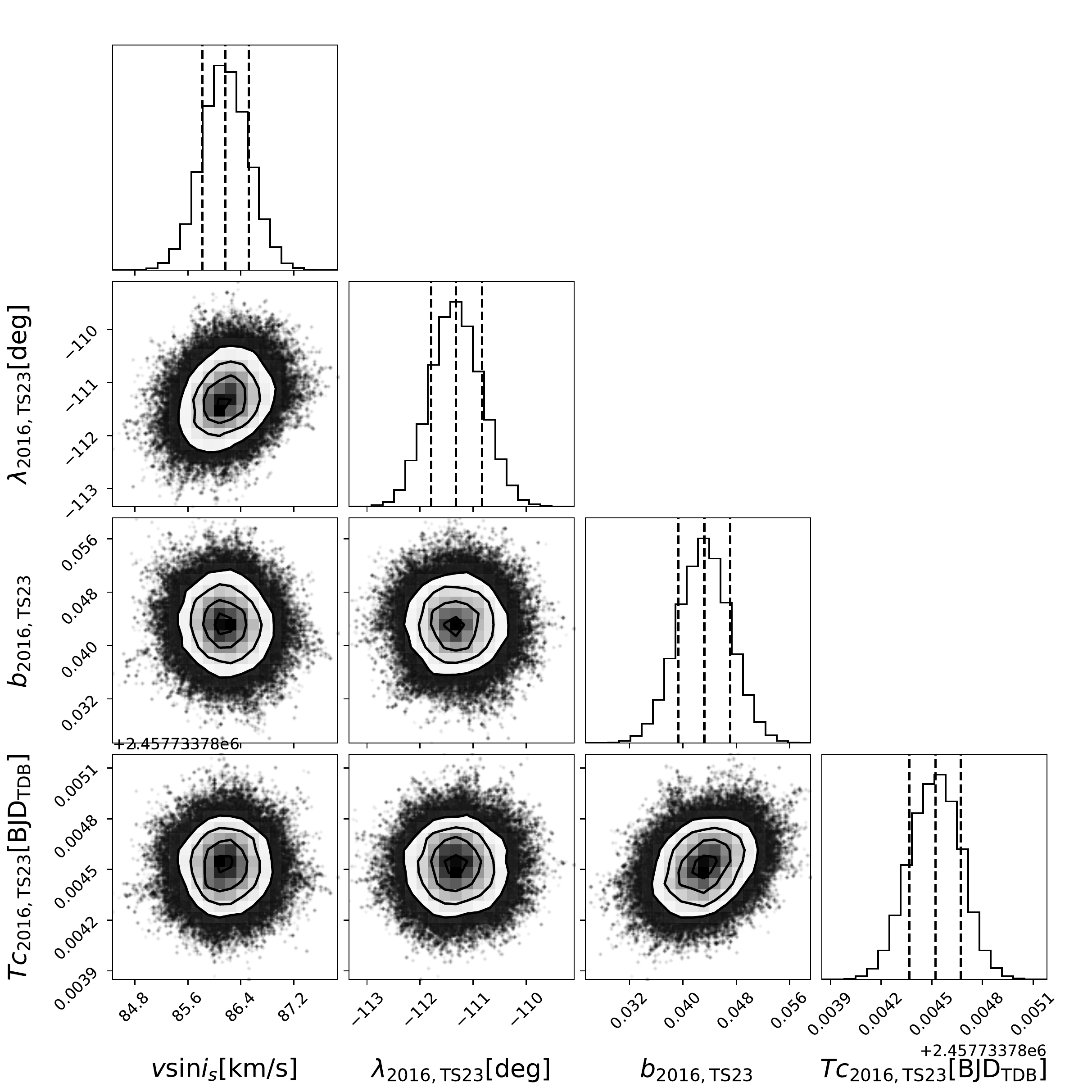}
 \end{minipage}
 \begin{minipage}{0.4\hsize}
   \includegraphics[width=75mm]{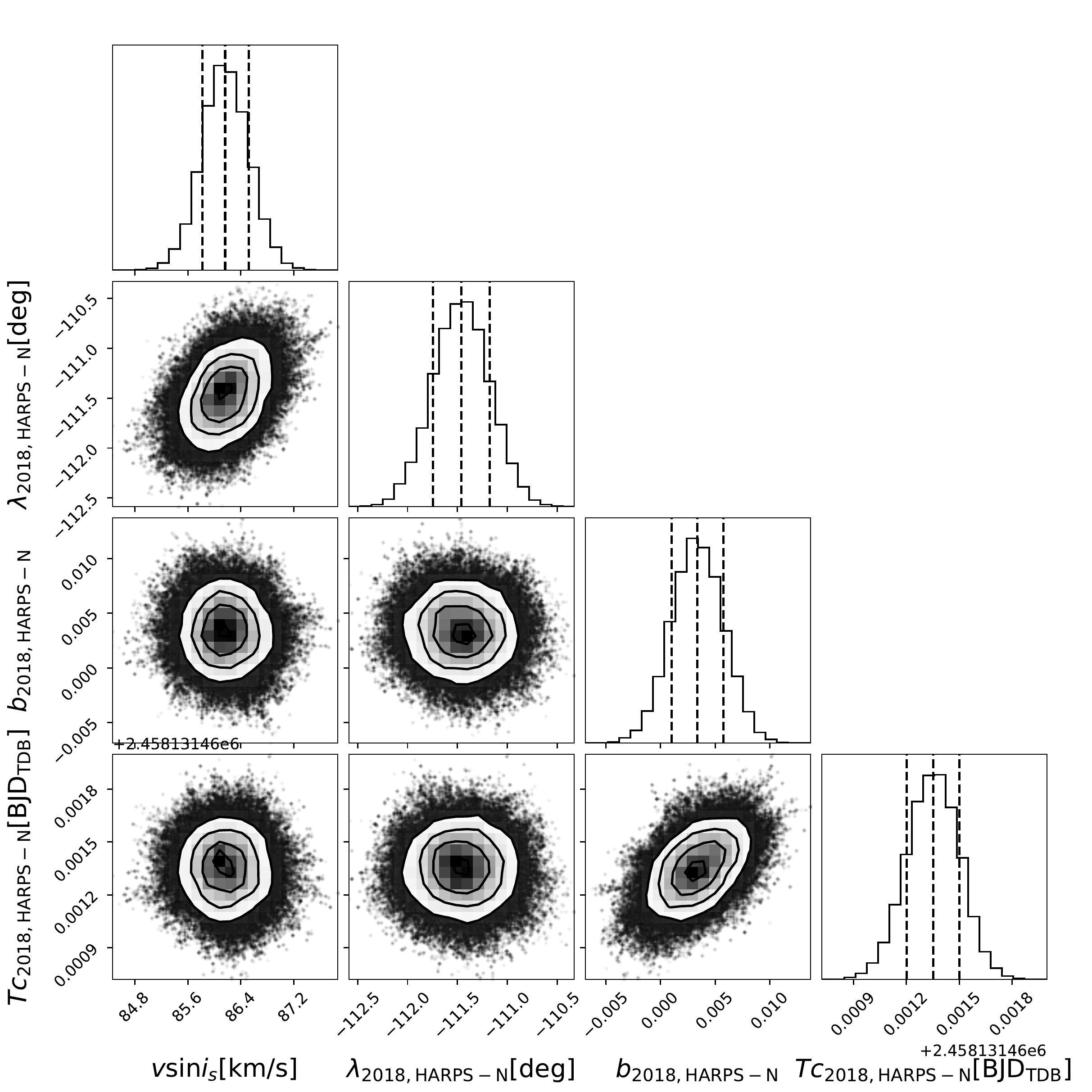}
 \end{minipage}
\caption{Corner plots for the free parameters after using MCMC in section \ref{DB}. Black circles indicate 68\%, 95\%, and 99.7\% confidence from the inside. In each posterior distribution of each parameter, the vertical dotted lines show the  best-fit value (middle) and 1$\sigma$ confidence (both ends). We created these plots using corner.py \citep{corner}.}
\label{figMCMC}
\end{figure*}

\begin{figure*}
\centering
\hspace*{-2cm}
\begin{minipage}{0.45\hsize}
   \includegraphics[width=75mm]{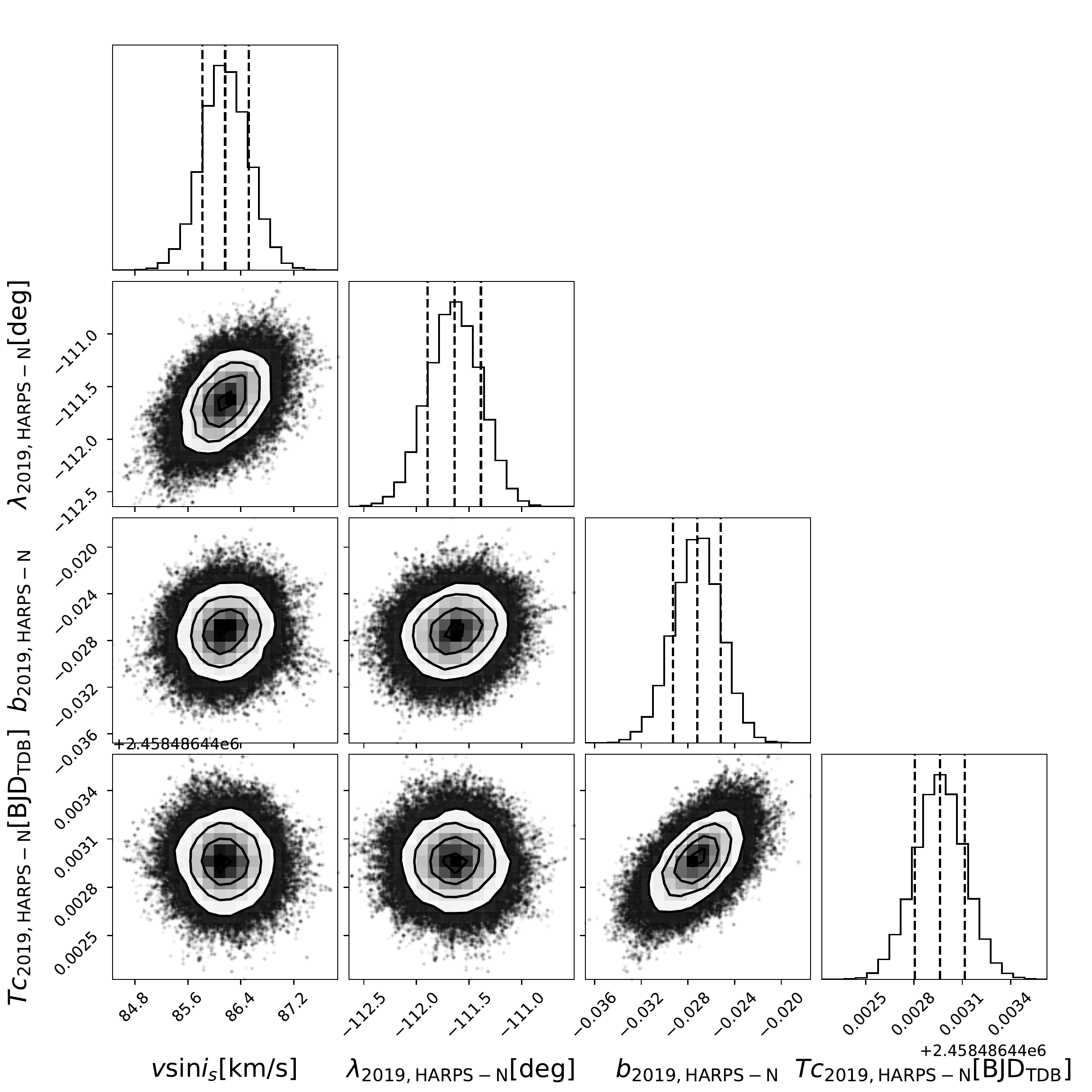}
 \end{minipage}
 \begin{minipage}{0.4\hsize}
   \includegraphics[width=75mm]{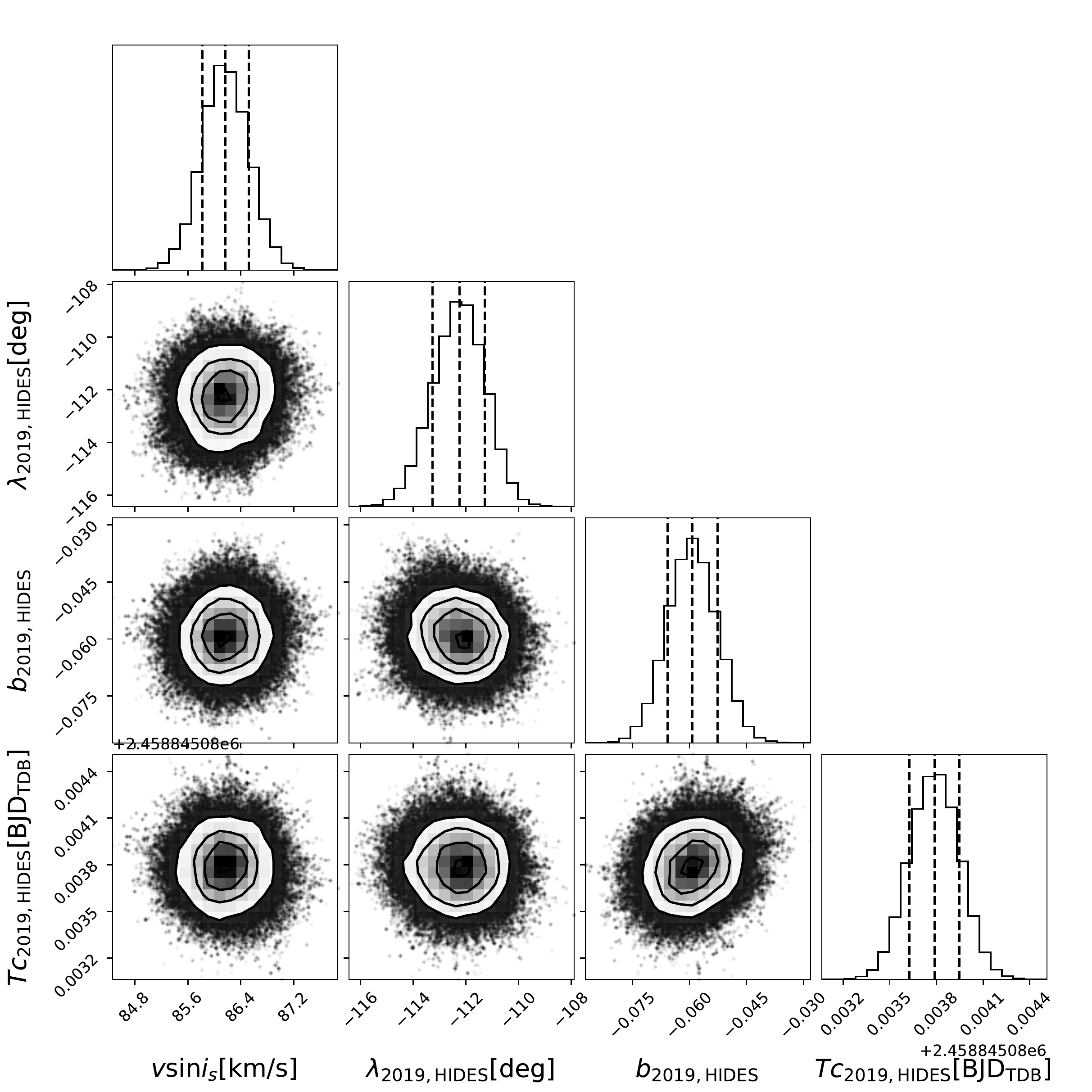}
 \end{minipage} \\
 
  \hspace*{-2cm}
 \begin{minipage}{0.45\hsize}
   \includegraphics[width=75mm]{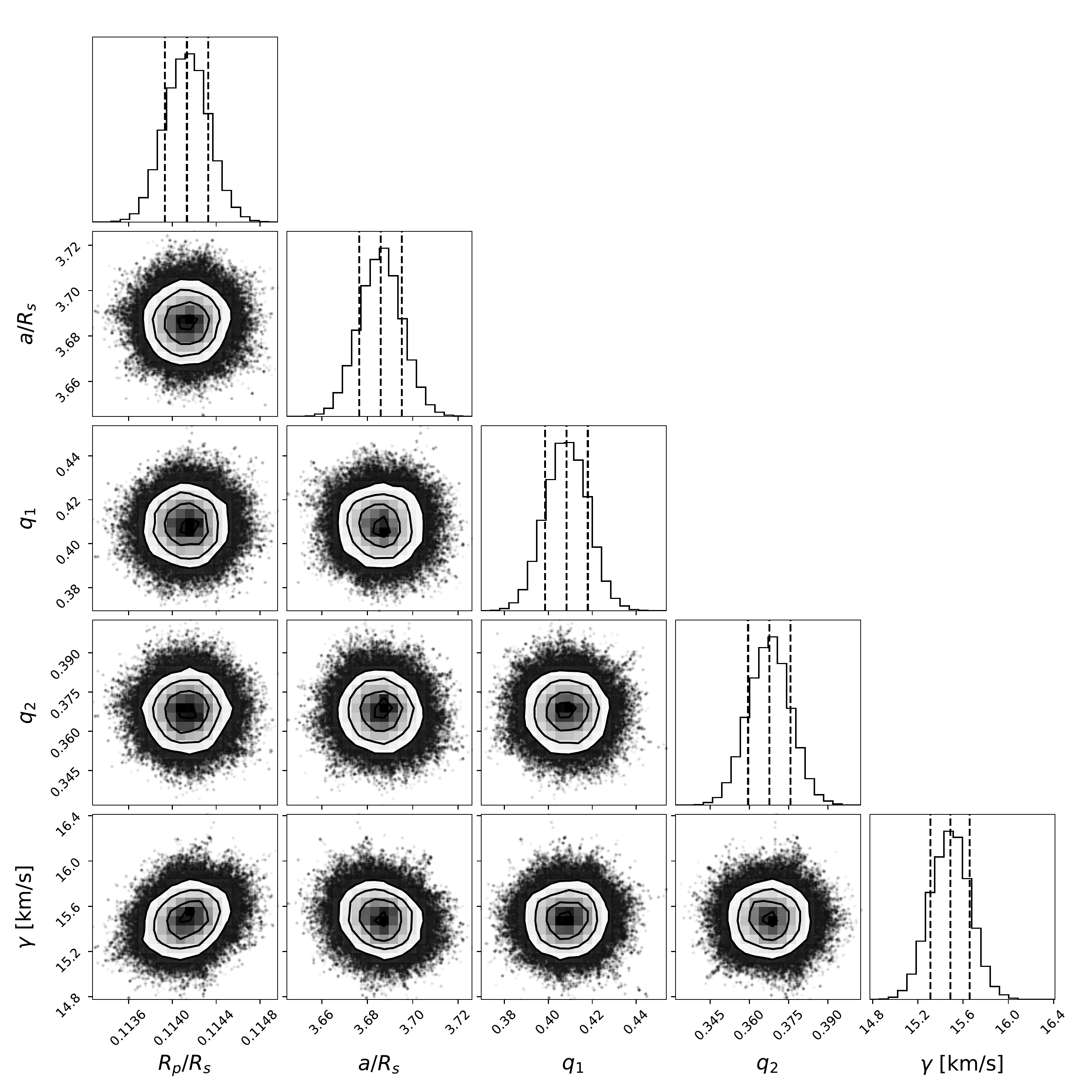}
 \end{minipage}
\caption{Continuance of Figure \ref{figMCMC}.}
\label{figMCMC_2}
\end{figure*}

\begin{figure*}
\centering
\begin{minipage}{0.45\hsize}
   \includegraphics[width=75mm]{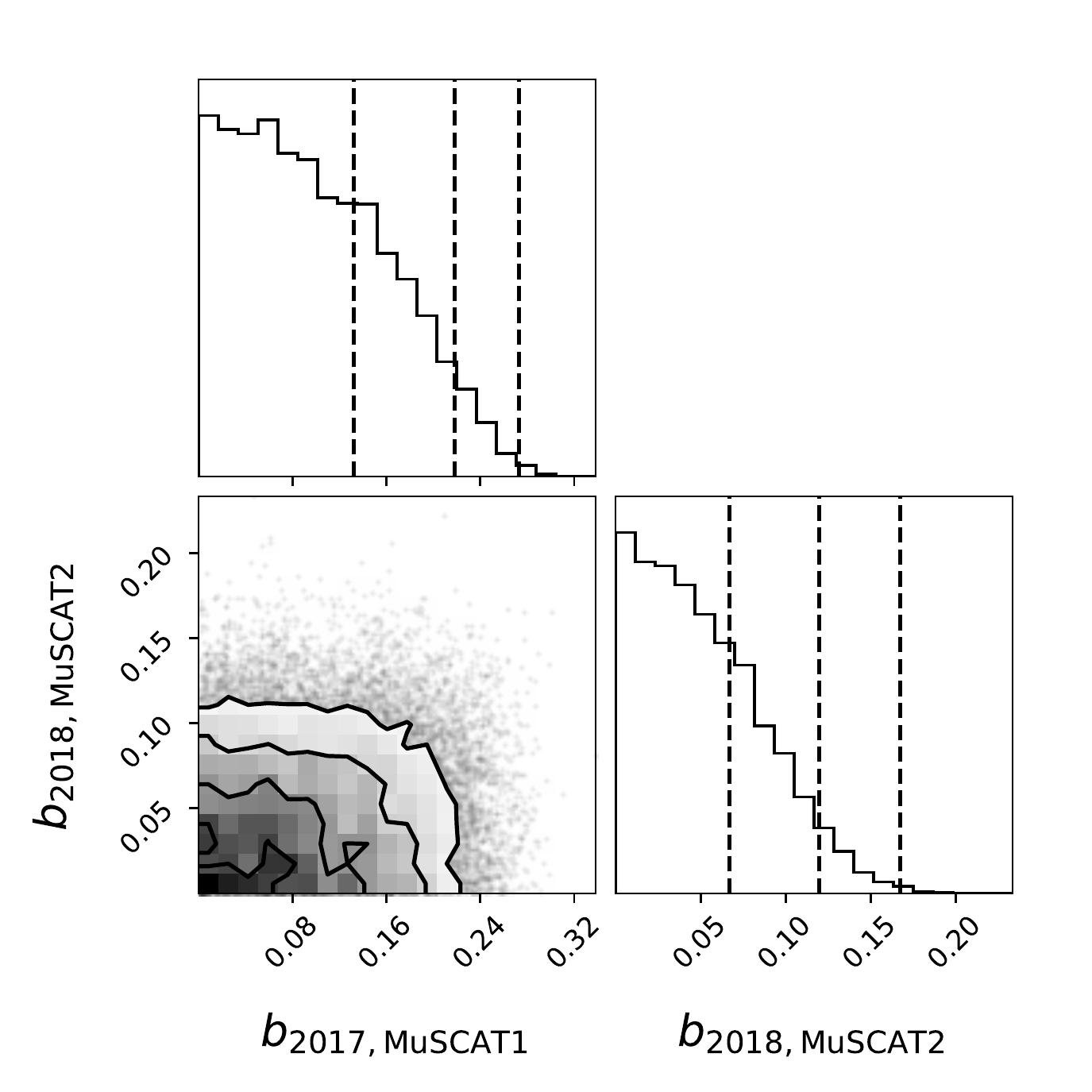}
 \end{minipage}
 \begin{minipage}{0.4\hsize}
   \includegraphics[width=75mm]{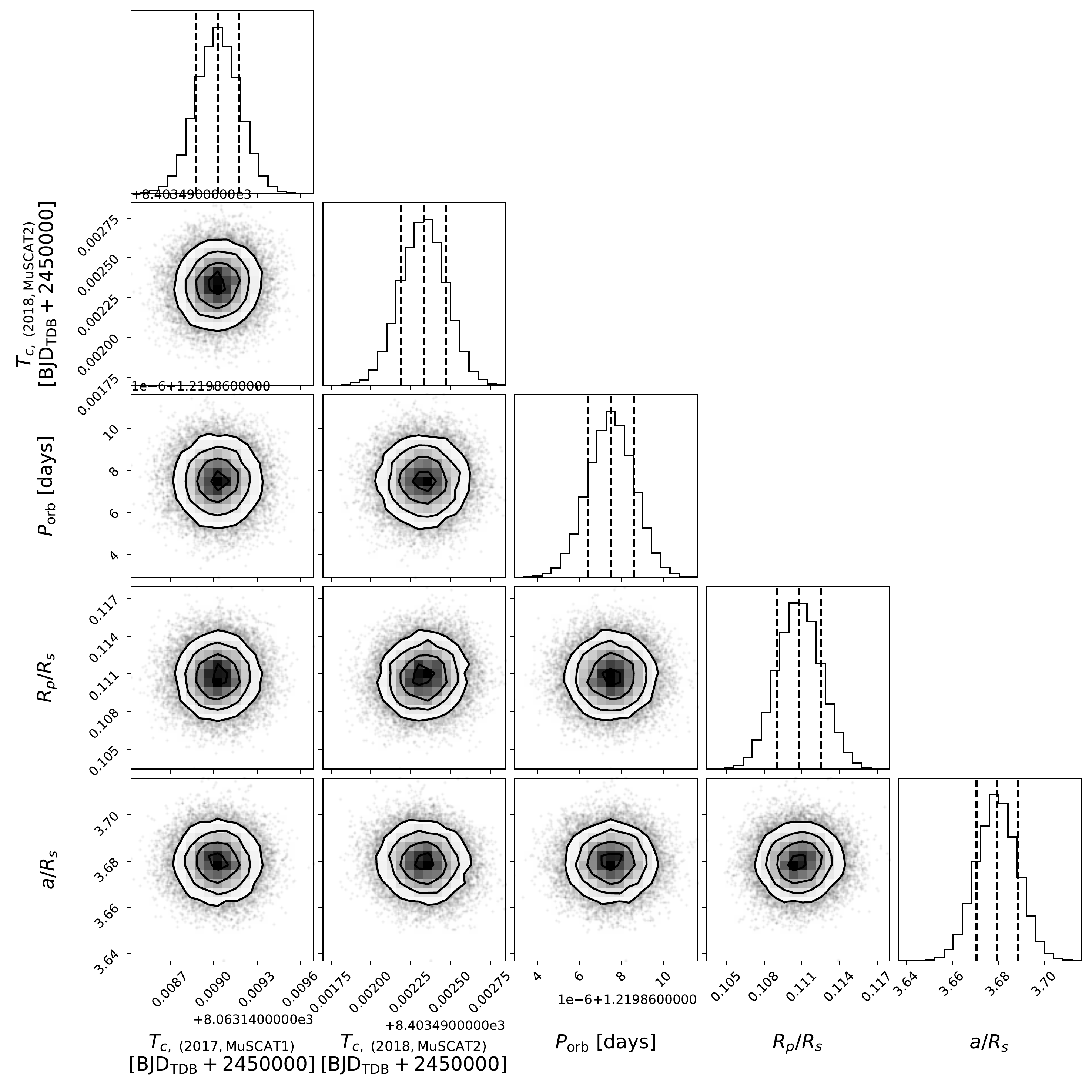}
 \end{minipage}\\
 \begin{minipage}{0.45\hsize}
   \includegraphics[width=75mm]{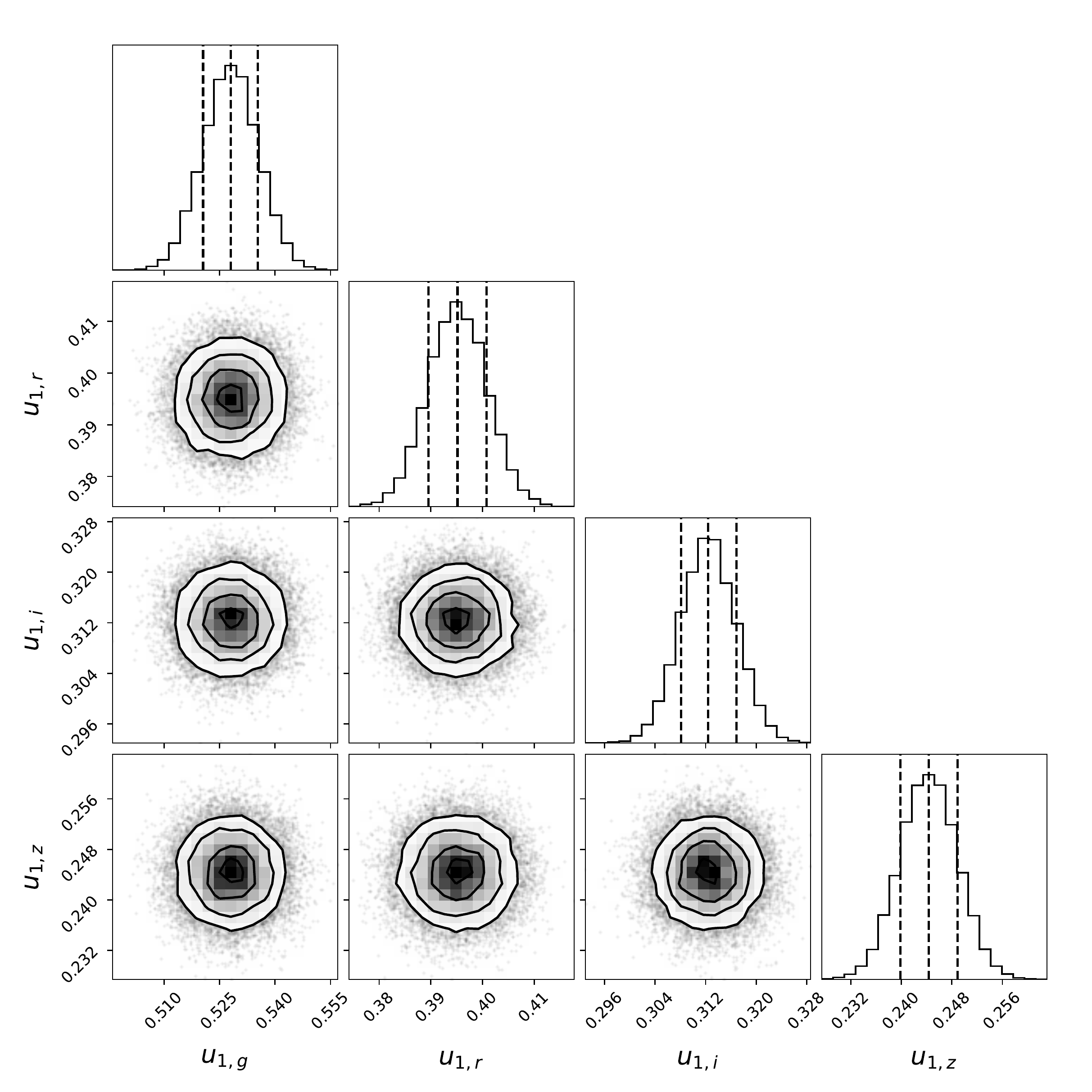}
 \end{minipage}
 \begin{minipage}{0.4\hsize}
   \includegraphics[width=75mm]{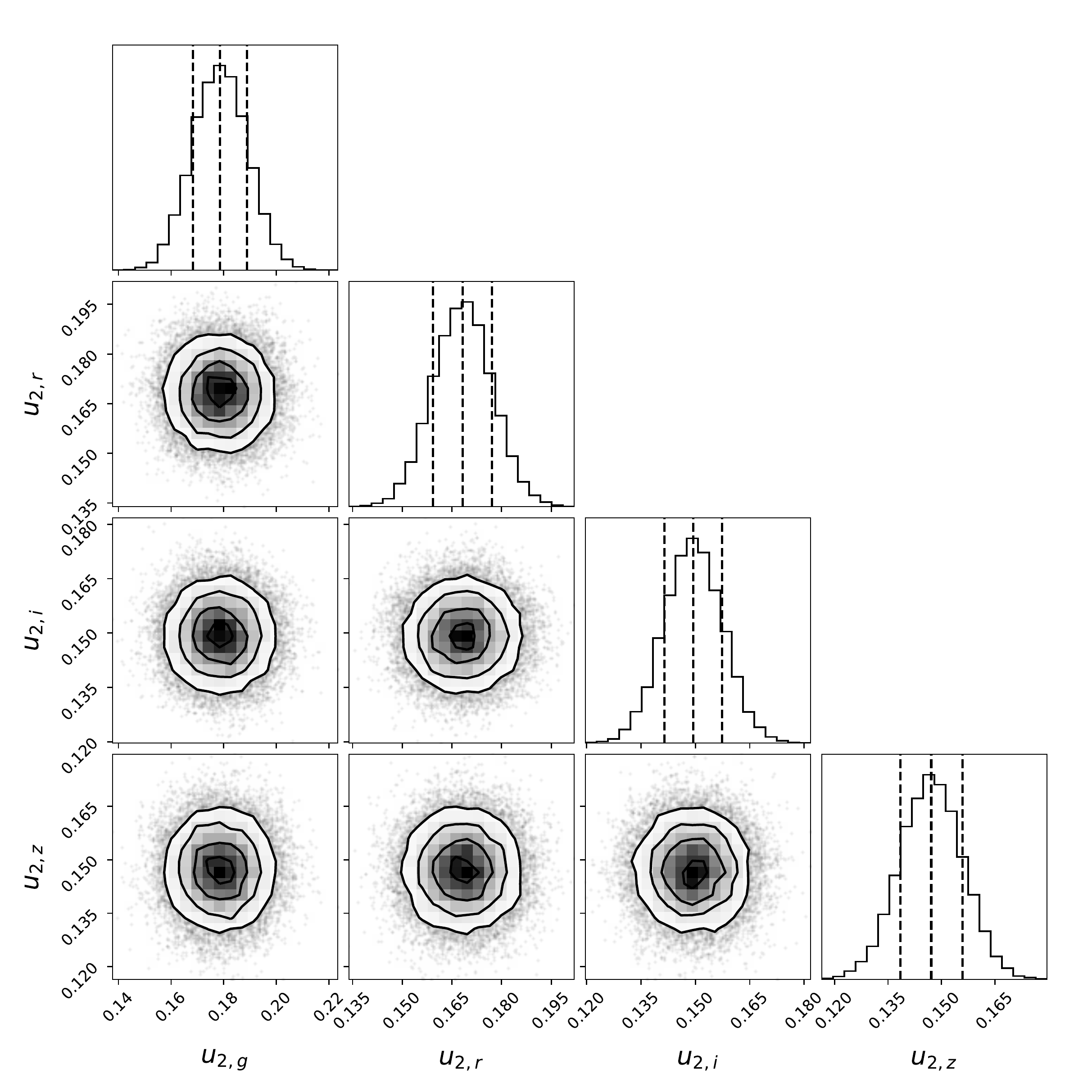}
 \end{minipage}\\
  \begin{minipage}{0.45\hsize}
   \includegraphics[width=75mm]{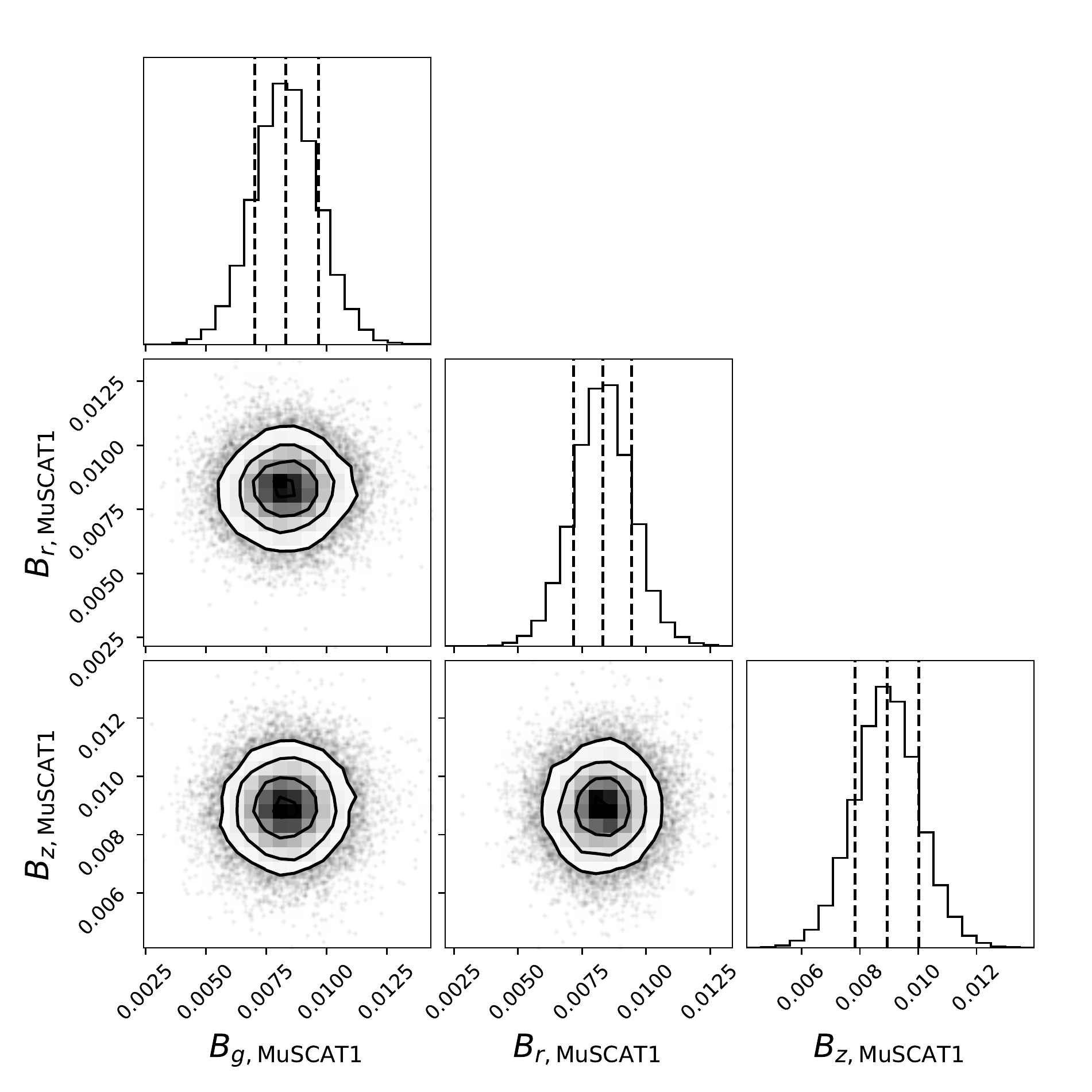}
 \end{minipage}
 \begin{minipage}{0.4\hsize}
   \includegraphics[width=75mm]{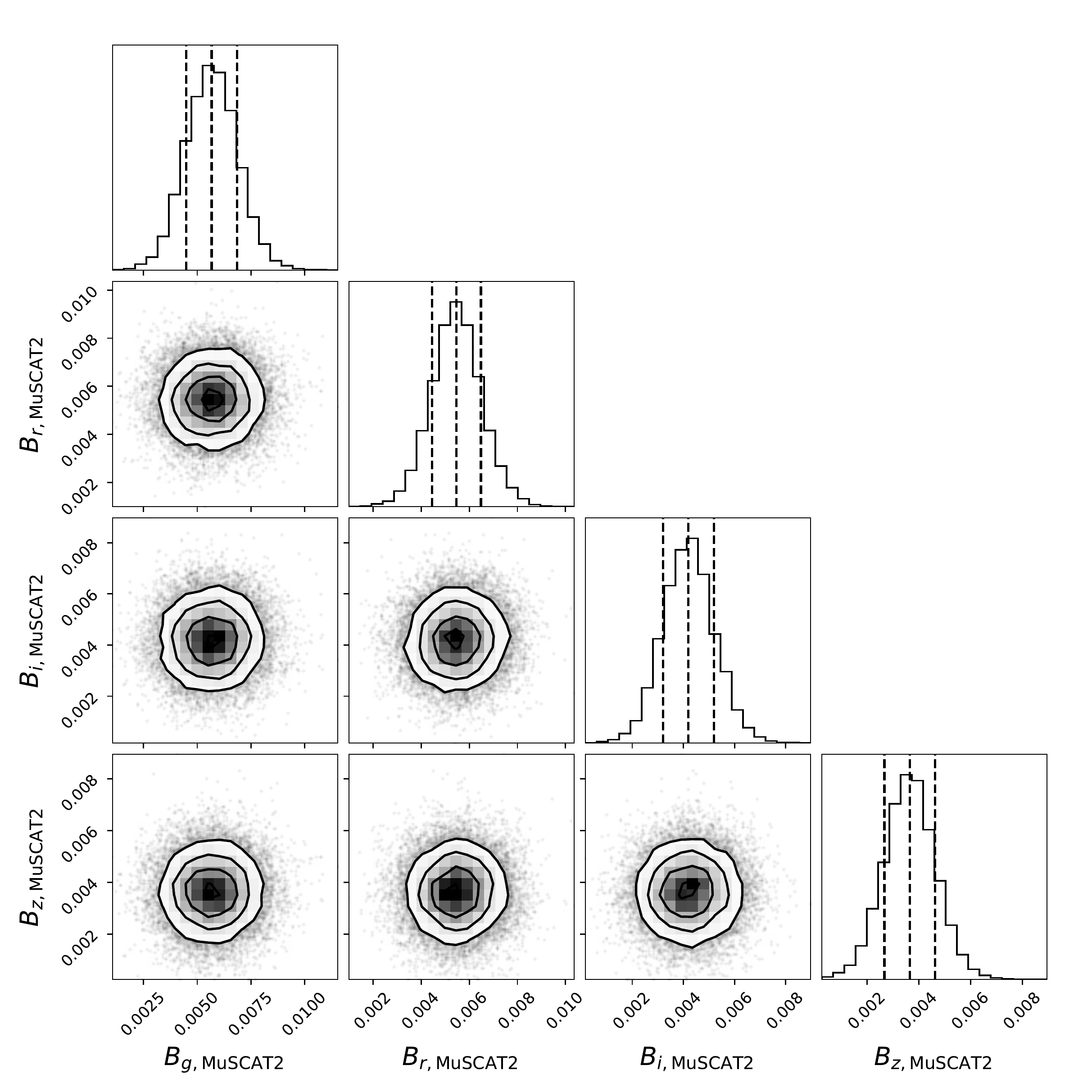}
 \end{minipage}\\
\caption{Plot distributions for impact parameter of each epoch, radial-ratio, period and transit mid-time of each epoch, limb darkening coefficients of each band, and base line of each light curve. These are the same corner plots as in Figure \ref{figMCMC}, but for the transit photometry by MuSCAT1 and MuSCAT2.}
\label{figMCMC_Ph}
\end{figure*}

\begin{figure*}[htbp]
\centering
\hspace*{-2cm}
\begin{minipage}{0.5\hsize}
   \includegraphics[width=75mm]{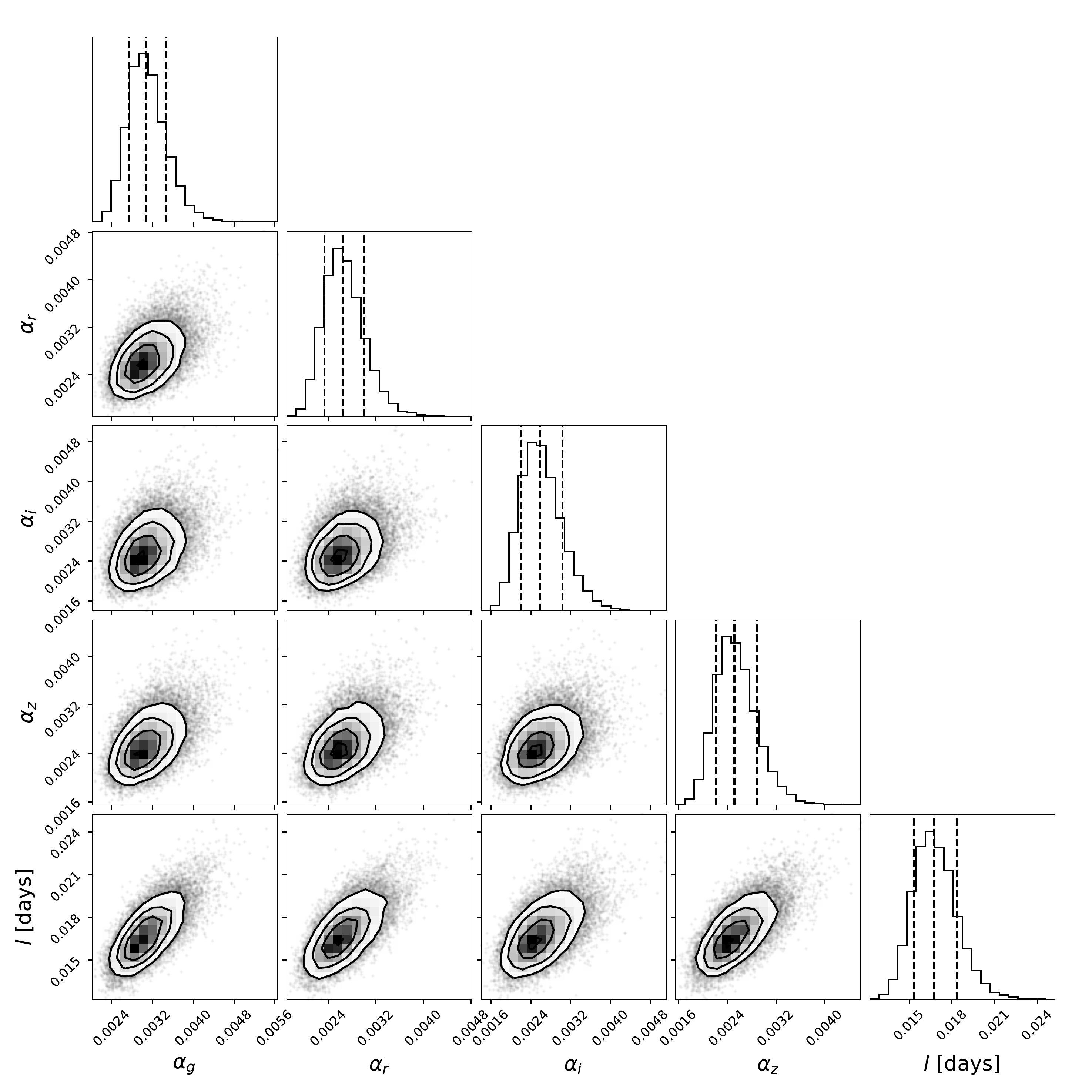}
 \end{minipage}\\
\caption{Continuance of Figure \ref{figMCMC_Ph}. Plot distributions for the impact parameter of each epoch, $\sigma$ of each band, $l$, period, and transit mid-time of each epoch.}
\label{figMCMC_Ph_2}
\end{figure*}

Here, we display the corner plots after using MCMC in Section \ref{DB} (Figure \ref{figMCMC} and Figure \ref{figMCMC_2}) and in Section \ref{LCF} (Figure \ref{figMCMC_Ph} and Figure \ref{figMCMC_Ph_2}).

\bsp 
\label{lastpage}
\end{document}